\documentclass[floats,floatfix,showpacs,amssymb,prd,twocolumn,superscriptaddress,nofootinbib,nolongbibliography,reprint]{revtex4-2}

\usepackage{amssymb,amsmath,verbatim,mathtools,needspace,enumitem,etoolbox,graphicx,physics,microtype,afterpage,bigints,gensymb,tabularx,xspace}

\usepackage[dvipsnames,table]{xcolor}
\definecolor{linkcolor}{rgb}{0.0,0.3,0.5}
\definecolor{dodgerblue}{HTML}{1E90FF}
\usepackage[unicode, colorlinks=true, linkcolor=linkcolor, citecolor=linkcolor, filecolor=linkcolor,urlcolor=linkcolor, pdfusetitle]{hyperref}
\usepackage[all]{hypcap}
\usepackage[T1]{fontenc}
\usepackage[utf8]{inputenc}
\usepackage{orcidlink}
\usepackage{lipsum}
\usepackage{booktabs}

\usepackage[normalem]{ulem}

\makeatletter
\def\tabular@hook{\col@sep\tabcolsep}
\makeatother

\makeatletter
\newcommand*{\balancecolsandclearpage}{\close@column@grid \cleardoublepage \twocolumngrid}
\makeatother

\newcommand{\milan}{\affiliation{Dipartimento di Fisica ``G. Occhialini'', Universit\'a degli Studi di Milano-Bicocca, Piazza della Scienza 3, 20126 Milano, Italy}}
\newcommand{\infn}{\affiliation{INFN, Sezione di Milano-Bicocca, Piazza della Scienza 3, 20126 Milano, Italy}}
\newcommand{\nottingham}{\affiliation{Nottingham Centre of Gravity \& School of Mathematical Sciences, University of Nottingham, University Park, Nottingham NG7 2RD, United Kingdom}}

\newcommand{\pOneG}[1]{\ensuremath{p_{\mathrm{1G}}\!\left(#1\right)}}

\newcommand{\ppop}{\ensuremath{p_{\rm pop}}\xspace}
\newcommand{\Nobs}{\ensuremath{N_{\rm obs}}\xspace}

\newcommand{\ppe}{\ensuremath{\pi}\xspace}

\newcommand{\model}[1]{\textsc{\small #1}\xspace}
\newcommand{\defaultBBH}{\model{Default BBH}}

\renewcommand{\det}{\ensuremath{{\rm det}}\xspace}

\usepackage{tikz}
\usetikzlibrary{arrows.meta,decorations.pathmorphing,calc,positioning,shadows.blur}

\begin{document}

\title{
Gravitational-wave hierarchical inference from population samples}
\author{Federico De Santi$\,$\orcidlink{0009-0000-2445-5729}}
\email{f.desanti@campus.unimib.it}
\milan
\infn

\author{Davide Gerosa$\,$\orcidlink{0000-0002-0933-3579}}
\milan
\infn

\author{Matthew Mould$\,$\orcidlink{0000-0001-5460-2910}}
\nottingham

\date{\today}

\begin{abstract}
Standard approaches to gravitational-wave population inference require evaluating
the probability density of the population model.
While appropriate for models composed of
exact functional forms,
this
approach
is not compatible with state-of-the-art
astrophysical predictions from population-synthesis codes,
which simulate samples from astrophysical models whose densities are unknown.
We present an approach
that solely requires samples and never density evaluations of the targeted population model.
Key technical ingredients include
efficient surrogates for the single-event likelihoods and detection probability,
together with an efficient sampling strategy.
Using detections up to GWTC-5.0, we validate the method against standard inference with a parametric model, showcase its flexibility on a simulated population including hierarchical mergers, and present a realistic application to a grid of astrophysical population-synthesis simulations with five formation channels.
This work paves the way toward a direct comparison of astrophysical models and gravitational-wave data.

\end{abstract}

\maketitle

\section{Introduction}\label{sec: introduction}

Population analyses are now among the most anticipated outcomes of gravitational-wave (GW) observing campaigns, with flagship LIGO--Virgo--KAGRA (LVK)~\cite{LIGOScientific:2014pky, VIRGO:2014yos, KAGRA:2020tym} studies~\cite{KAGRA:2021duu,LIGOScientific:2025pvj,LIGOScientific:2026ctl} stimulating a wealth of further analyses~\cite{CALLISTER2026546}.
The distributions of black hole (BH) masses, spins, and redshifts probe the processes that assemble merging binaries, including isolated stellar evolution and dynamical encounters in dense star clusters~\cite{Mandel:2018hfr,Mapelli:2021taw}.
Hierarchical Bayesian inference reconstructs these distributions by combining individual event likelihoods with
detector selection effects~\cite{Mandel:2018mve,Vitale:2020aaz}.

In standard implementations, hierarchical inference requires the density of the population model to be evaluated at arbitrary points in parameter space.
This restricts those models to analytical closed-form densities,
either parametric (e.g.,~\cite{KAGRA:2021duu,LIGOScientific:2025pvj,LIGOScientific:2026ctl})
or nonparametric (e.g.,~\cite{Edelman:2022ydv, Callister:2023tgi, Ray:2023upk, Heinzel:2024jlc, Rinaldi:2021bhm, Tenorio:2025nyt}).
In both cases, the connection with astrophysics is drawn after the inference, by comparing the measured features in the population-level distribution, such as peaks,
breaks, and cutoffs, with the predictions of formation scenarios. 
This ``comparison'' is, at present, still performed somewhat loosely, often by simply overplotting distributions.

Population-synthesis codes (e.g., for isolated binaries~\cite{Spera:2015vkd, Breivik:2019lmt, COMPASTeam:2021tbl, Fragos:2022jik} and dense star clusters~\cite{Rodriguez:2021qhl,Kritos:2022ggc}), on the
other hand, directly encode the astrophysics driving binary BH
mergers and represent our best attempt to predict the population from first principles.
State-of-the-art codes evolve a set of initial conditions  and return a finite catalog of merging binaries.
The population is defined implicitly by the simulation code, and its density is not available in closed form. Such outputs cannot be used directly in the standard framework for hierarchical inference.
 In statistical terms, we have access to a forward model, while inference requires solving the inverse problem.
In practice, astrophysical simulations provide a population model that is ``easy to sample'' (by running the simulation) but ``hard to evaluate'' (by computing its probability density at a given point).
This is a central obstacle to comparing astrophysical models with GW data, leaving such comparisons at a somewhat heuristic level.

Previous studies have addressed this issue by replacing the population model with an intermediate density estimate constructed from the simulated binaries.
This involves running population-synthesis simulations on a grid and interpolating the conditional probability density of the BH parameters across the population hyperparameter space.
A variety of density estimation and interpolation methods have been used, including kernel density estimation~\cite{Zevin:2020gbd,Cheng:2023ddt}, Gaussian process regression~\cite{Barrett:2016edh,Taylor:2018iat}, neural networks~\cite{Mould:2022ccw,Riley:2023jep}, and normalizing flows~\cite{Wong:2020ise,Colloms:2025hib,Plunkett:2025mjr}.

Employing intermediate density estimators fitted to simulations has several drawbacks,
however.
First, uneven coverage of the population parameter space and varying numbers of binaries across simulations can introduce systematic uncertainties into the population inference that are difficult to quantify.
On the practical side, these estimators must be rebuilt and validated for each new set of simulations, whenever population-synthesis predictions are revised or the input physics is updated.
Simulation-based inference offers an alternative that avoids these density evaluations~\cite{Cranmer:2019eaq,Leyde:2023iof,Leyde:2026hvm}, but likewise requires retraining a neural network for each new set of simulations.
Another strategy separates the description of the data from their astrophysical interpretation by deriving astrophysical results from existing, ideally nonparametric GW population fits~\cite{Wong:2022bxp,Fabbri:2025faf,Rinaldi:2025evs}.
However, this approach assumes an ad hoc conversion from existing Bayesian fits to the population.
\clearpage
In this paper, we instead tackle the problem the other way around, and develop a general framework
to perform hierarchical inference by sampling the population model instead of evaluating it (Sec.~\ref{sec: strategies}).
Our method is tailored to the general case where the population model is only available through samples and does not require interpolations of the population density or analytical approximations of the likelihood.
Specifically, we rewrite
the population likelihood so that samples are drawn from the population model, while the functions to be evaluated are the single-event likelihoods and detection probability,
which thus require efficient surrogates (Sec.~\ref{sec: methods}).

Such approaches have been individually explored in the literature.
For the contributions from individual events, surrogates have been fitted to their likelihoods or reference posterior samples using
(multivariate) Gaussian approximations
\cite{
Delfavero:2023nbl, Qazalbash:2026wyw, Mould:2023eca},
kernel density estimation (KDE)~\cite{Callister:2022qwb},
Gaussian processes~\cite{DEmilio:2021laf},
Gaussian mixture models (GMMs)~\cite{Rinaldi:2021bhm, Golomb:2021tll, Mancarella:2025uat, DeRenzis:2026hcp, Hussain:2025llf}, 
or more generic proposal distributions used for importance sampling~\cite{Yeung:2026dne};
alternatively, generative models like normalizing flows can draw extra parameter estimation (PE) samples after fitting to an existing sample set~\cite{Adamcewicz:2026pai}, while simulation-based inference approaches instead train a posterior estimator directly~\cite{Green:2020dnx, Dax:2021tsq}.
Surrogates to estimate the contribution due to selection effects have been made using neural networks~\cite{Gerosa:2020pgy, Callister:2024qyq, Gassert:2026ugv},
physically motivated semianalytic models~\cite{Essick:2023toz,Mould:2023eca,Gerosa:2024isl} or fitted parametrized models~\cite{Lorenzo-Medina:2024opt},
and GMMs fitted to a reference detectable population~\cite{Talbot:2020oeu}.

Here, we build upon and extend these approaches to enable evaluations of the population likelihood using population simulations, using GMMs for the single-event likelihoods and detection probability. A working implementation further requires a novel strategy to restrict population samples to regions of parameter space where these functions each appreciably contribute to the overall likelihood; we develop such a strategy to prevent the inference from being dominated by the variance of Monte Carlo estimators~\cite{Farr:2019rap,Essick:2022ojx,Talbot:2023pex,Heinzel:2025ogf,Mancarella:2025uat}. For selection effects, we additionally generalize the GMM approach to account for multiple observing runs with varying sensitivity, durations, and injection distributions.

We apply our framework to three cases of increasing complexity,
all using data from the latest cumulative GW catalog, GWTC-5.0~\cite{LIGOScientific:2018mvr,LIGOScientific:2020ibl,LIGOScientific:2021usb,KAGRA:2021vkt,LIGOScientific:2025slb,LIGOScientific:2026wfs}.

\begin{enumerate}
\item
We first validate our procedure against standard inference using a parametric model adopted by the LVK Collaboration~\cite{LIGOScientific:2025pvj, LIGOScientific:2026ctl} (Sec.~\ref{sec: comparison}).
For this test where the population model can both be sampled and evaluated, we show that our approach yields posteriors consistent with the standard results, thus confirming the robustness of the method.

\item Second, we consider a parametric model augmented with hierarchical mergers~\cite{Gerosa:2021mno} (Sec.~\ref{sec: hierarchical mergers}).
Remnants of mergers between BHs in an underlying ``first-generation'' population form a ``second-generation'' population whose members are allowed to merge again.
The remnant properties are predicted using fits to numerical-relativity simulations. These fits are nonlinear functions of the progenitor properties and generally do not admit a unique inverse.
Hierarchical mergers therefore provide an ideal test case in which only the forward model is available: sampling the population amounts to evaluating these remnant predictions, while the population density itself is not explicitly known.

\item Finally, we present a realistic application to population-synthesis predictions using publicly available simulations from Ref.~\cite{Zevin:2020gbd}, which describe five formation channels on a grid of natal BH spins and common-envelope efficiencies (Sec.~\ref{sec: pop synth}).
Our method allows us to evaluate the likelihood directly using the simulated binaries and consistently rank the simulated scenarios, without requiring density estimation or interpolation of the astrophysical simulations themselves.

\end{enumerate}

This study thus addresses the longstanding challenge of a direct comparison between astrophysical models and GW catalogs (Sec.~\ref{sec: conclusions}).

\section{Two possible strategies}
\label{sec: strategies}

\subsection{Population likelihood}

Assume that data $\mathcal{D} = \{d_i\}_{i=1}^{\Nobs}$ are available for a GW catalog of $\Nobs$ observed events. For each individual event, PE has been performed, yielding posterior samples of the source parameters, $\theta\sim p(\theta|d_i)$.
The goal of population inference is to infer the distribution of source parameters generating the observed events, namely
the population model $\ppop(\theta|\lambda)$, where $\lambda$ are the hyperparameters.

The likelihood $p(\mathcal{D}|\lambda)$ is that of an inhomogeneous Poisson process with selection effects~\cite{Mandel:2018mve, Vitale:2020aaz}, given by
\begin{equation}\label{eq: standard hierarchical likelihood}
    \begin{split}
    p(\mathcal{D}|\lambda) &= {\xi(\lambda)^{-\Nobs}}\prod_{i=1}^{\Nobs} p(d_i|\lambda)\,,\\
    \end{split}
\end{equation}
where $p(d_i|\lambda)$ is the likelihood of the $i$-th observation and  $\xi(\lambda)$ is the selection function.  The likelihood contribution of each observation is expressed in terms of the population model $\ppop(\theta|\lambda)$ and the individual-event likelihood $p(d_i|\theta)$ as
\begin{equation}\label{eq: standard hierarchical likelihood event}
    \begin{split}
    p(d_i|\lambda) &= \int \dd{\theta}\,p(d_i|\theta) \ppop(\theta|\lambda)
\propto \int \dd{\theta}\,\frac{p(\theta|d_i)}{\ppe(\theta)} \ppop(\theta|\lambda)\;,
    \end{split}
\end{equation}
where we have used Bayes' theorem to express $p(d_i|\theta)$ as proportional to the ratio of the posterior $p(\theta|d_i)$ to the PE prior $\ppe(\theta)$.
To simplify the notation, we henceforth denote $p(\mathcal{D}|\lambda)$ by $\mathcal{L}(\lambda)$. Selection effects are encoded in the term
\begin{equation}\label{eq: selection effects}
    \xi(\lambda)
    = \int \dd{\theta}\,p(\det|\theta)\,\ppop(\theta|\lambda) \,,
\end{equation}
where $p(\det|\theta) \in [0,1]$ is the probability of detecting a source with parameters $\theta$, marginalized over data realizations.

The population model $\ppop$ appears in two places in the likelihood of Eq.~\eqref{eq: standard hierarchical likelihood}: in the product over individual-event likelihoods [cf. Eq.~\eqref{eq: standard hierarchical likelihood event}] and in the selection function [cf. Eq.~\eqref{eq: selection effects}].
In both cases, the integrals in $\mathcal{L}(\lambda)$ are
typically computed using Monte Carlo integration. We consider two estimators: one based on evaluating the  density of the population model, as in the standard approach, and the other based on drawing samples from it, as in the new approach introduced in this work.

\subsection{The standard approach: evaluating $\ppop(\theta|\lambda)$}
If the population model density $\ppop(\theta|\lambda)$ can be evaluated at arbitrary points in parameter space, the likelihood in Eq.~\eqref{eq: standard hierarchical likelihood} can be estimated as
\begin{align}    \hat{\mathcal{L}}(\lambda) &= \hat{\xi}_{\rm inj}(\lambda)^{-\Nobs}\prod_{i=1}^{\Nobs} \left[\frac{1}{N_{{\rm s}, i}}\sum_{k=1}^{N_{{\rm s}, i}}\frac{\ppop(\theta_{ki}|\lambda)}{\ppe(\theta_{ki})}\right]\,,
\notag \\
\theta_{ki} &\sim p(\theta|d_i)\,,
\label{eq: standard monte carlo likelihood estimator}
\end{align}
where we have explicitly written the Monte Carlo average over the $N_{{\rm s}, i}$ posterior samples $\theta_{ki}$ for the $i$-th event.
Similarly, the selection function is estimated by importance reweighting a set of injections drawn from a reference distribution $q(\theta)$, of which $N_{\rm found}$ pass a detection threshold~\cite{Tiwari:2017ndi}.
The resulting estimator is
\begin{align}
    \hat{\xi}_{\rm inj}(\lambda) &= \frac{1}{N_\mathrm{inj}} \sum_{k=1}^{N_\mathrm{found}} w_k \,\frac{\ppop(\theta_k|\lambda)}{q(\theta_k)}\,,
   \quad \theta_k \sim q(\theta)\,,
\label{eq: Monte Carlo Pdet estimator}
\end{align}
where $N_{\rm inj}$ is the total number of injections (found and missed). The weights $w_k$ ensure that injections from different observing runs are combined appropriately, accounting for differences in run duration and injection distributions~\cite{Essick:2025zed}. Both $w_k$ and $q(\theta_k)$ are known for each injection and are provided in the public data releases~\cite{LIGOScientific:2026jgl}. 

For the standard approach, the main ingredients can therefore be summarized as follows:
\begin{enumerate}
\item A population model $\ppop(\theta|\lambda)$ that can be evaluated at arbitrary points in the hyperparameter space.
\item Samples from individual-event posteriors $p(\theta|d_i)$.
\item A set of injections which are reweighted to the population model $\ppop(\theta|\lambda)$ to estimate the selection function $\xi(\lambda)$.
\end{enumerate}

\subsection{An alternative approach: sampling from $\ppop(\theta|\lambda)$}
Let us now consider the case where the probability density function $\ppop(\theta|\lambda)$ is difficult to evaluate but can be sampled from more easily.
We rewrite the likelihood in Eq.~\eqref{eq: standard hierarchical likelihood} so that the Monte Carlo averages can be computed using only samples drawn from the population model itself.
In particular,
\begin{align}
\hat{\mathcal{L}}(\lambda) &= \hat{\xi}(\lambda)^{-\Nobs}\prod_{i=1}^{\Nobs} \left[\frac{1}{N_{\lambda, i}}\sum_{k=1}^{N_{\lambda, i}}\frac{p(\theta_{ki}|d_i)}{\ppe(\theta_{ki})}\right]\,,
\notag \\
\theta_{ki} &\sim \ppop(\theta|\lambda)\,,
\label{eq: sampling p_pop monte carlo likelihood estimator}
\end{align}
where a separate set of $N_{\lambda,i}$ population samples is drawn for each event $i$. The event index in $\theta_{ki}$ makes this distinction explicit: different events use different sets of independent samples, all drawn from the same population model $\ppop(\theta|\lambda)$.

Compared to the standard estimator in Eq.~\eqref{eq: standard monte carlo likelihood estimator}, evaluating Eq.~\eqref{eq: sampling p_pop monte carlo likelihood estimator} requires the following ingredients:
\begin{enumerate}
\item An unnormalized function proportional to each of the single-event likelihoods $p(d_i|\theta) \propto p(\theta|d_i)/\pi(\theta)$.
\item An estimator of the selection function $\xi(\lambda)$ that uses samples drawn from the population model $\ppop(\theta|\lambda)$ without reweighting injections.
\item A method for efficiently drawing population samples within the support of each individual-event posterior $p(\theta|d_i)$ to reduce the Monte Carlo variance~\cite{Heinzel:2025ogf}. This is motivated by the fact that individual-event posteriors typically have much narrower support than the population model.
\end{enumerate}
We now describe in detail how we address each of these points in our implementation.

\section{Sampling instead of evaluating}\label{sec: methods}

\subsection{Evaluating the event likelihoods}\label{sec: gmm fit posteriors}
\begin{figure*}[]
    \centering
    \includegraphics[width=0.69\textwidth]{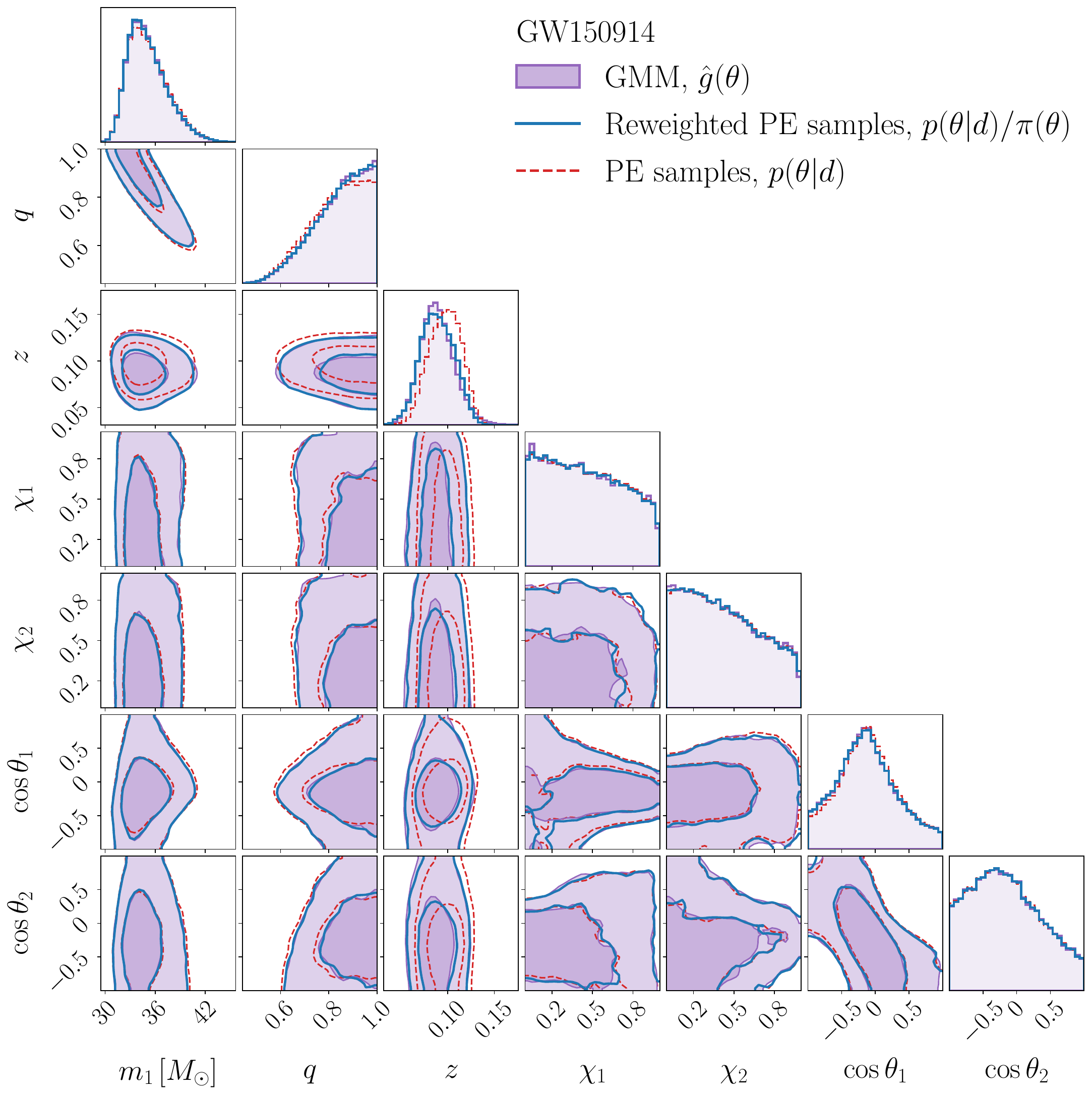}
    \caption{GMM fit to the posterior of GW150914.
    Draws from the GMM $\hat{g}_i(\theta)$ (purple filled) are compared with the posterior PE samples weighted by the inverse of the prior (blue solid),
    which are the target of the fit, as well as the unweighted PE samples (red dashed) for reference.
    Contours enclose the 50\% and 90\% credible regions. Masses are in the source frame.
    }
    \label{fig: gmm fit}
\end{figure*}

To evaluate the individual event likelihoods in Eq.~\eqref{eq: sampling p_pop monte carlo likelihood estimator},
it is sufficient to have an unnormalized function proportional to each $p(d_i|\theta)$.
In principle, one could use the PE likelihood itself~\cite{Thrane_2019}, but this is often computationally expensive due to the cost of waveform generation.
Instead, we construct a density estimate of the events' posterior-to-prior ratios $p(\theta|d_i)/\ppe(\theta)$, by means of a Gaussian mixture model,
using the PE samples publicly released by the LVK Collaboration.\footnote{We consider the 104 O4b candidates from the GWTC-5.0 data release~\cite{ligo_scientific_collaboration_and_virgo_2026_20348005, ligo_scientific_collaboration_and_virgo_2026_20348006},
the 86 O4a candidates from the GWTC-4.1 reanalysis~\cite{ligo_scientific_collaboration_and_virgo_2026_20275769, LIGOScientific:2026wfs},
and the 23 O3b candidates from GWTC-3~\cite{ligo_scientific_collaboration_and_virgo_2023_8177023, KAGRA:2021vkt}.
Of the 46 O1, O2, and O3a candidates, the 20 with NRSur7dq4 samples are taken from the GWTC-2 release~\cite{GWTC2_PE_DR, LIGOScientific:2020ibl},
and the remaining 26 from GWTC-2.1~\cite{ligo_scientific_collaboration_and_virgo_2022_6513631, LIGOScientific:2021usb}.
For each event, we use samples obtained with the NRSur7dq4 waveform model~\cite{Varma:2019csw} when available; otherwise, we concatenate samples obtained with the other available waveform models, typically  IMRPhenomXPHM~\cite{Pratten:2020ceb} and SEOBNRv5PHM~\cite{Ramos-Buades:2023ehm} or SEOBNRv4PHM~\cite{Ossokine:2020kjp}.}

For each event, we consider posterior samples over the following parameters:
\begin{equation}\label{eq: posterior parameters}
    \theta=\{m_1, q, z, \chi_1, \chi_2, \cos\theta_1, \cos\theta_2\}\,,
\end{equation}
namely, the source-frame primary mass $m_1$, the mass ratio $q=m_2/m_1$, the merger redshift $z$,
the dimensionless spin magnitudes $\chi_{1,2}$, and the cosines $\cos\theta_{1,2}$ of the tilt angles between
each spin and the orbital angular momentum.
Throughout this work, we assume that the population model factorizes into a distribution over the parameters $\theta$ of Eq.~\eqref{eq: posterior parameters} and one over the remaining binary parameters (e.g., sky position, binary orientation, and spin azimuthal angles).
For the latter, we implicitly adopt the distribution used both as the PE prior and to draw the injections, which is the same for these parameters.

Samples are provided in terms of detector-frame masses and luminosity distance.
We convert luminosity distance $d_L$ to redshift $z$ assuming the Planck 2018 cosmology~\cite{Planck:2018vyg}
and obtain the source-frame primary mass as $m_1=m_1^{\rm det}/(1+z)$.
The PE prior is uniform in the two detector-frame component masses, the spin magnitudes, and the cosines of the tilt angles.
Therefore,
\begin{equation}\label{eq: pe prior fit coordinates}
    \ppe(\theta) \propto \pi(d_L) \left|\frac{\dd{d_L}}{\dd{z}}\right| m_1 (1+z)^2
    \,,
\end{equation}
where $\pi(d_L)$ is the distance prior used in the corresponding catalog release. The factor $\mathrm{d}d_L/\mathrm{d}z$ converts this prior into a density in redshift, while $m_1(1+z)^2$ accounts for the change of variables from $(m_1^\det, m_2^\det)$ to $(m_1, q)$.
We then weight each PE sample $\theta_k$ by $1/\ppe(\theta_k)$
and fit a GMM to the weighted samples using an expectation–maximization (EM) algorithm.

A GMM has support over the entire 7-dimensional parameter space $\mathbb{R}^P$, whereas our parameters $\theta$ satisfy the constraints $m_1>0$, $q\in(0,1]$, $z>0$, $\chi_{1,2}\in[0,1]$, and $\cos\theta_{1,2}\in[-1,1]$.
Fitting a mixture directly in this constrained parameter space would require a very large number of components to accurately capture boundary effects.
Reference~\cite{Hussain:2025llf} addresses this by truncating each component.
We instead map each parameter to an unbounded coordinate $y_p=T(\theta_p)$ and fit a mixture with
full covariance matrices in these coordinates, similar to Ref.~\cite{Mancarella:2025uat}.

For the mass parameters and redshift, we use
\begin{equation}\label{eq: analytic maps}
    y_{m_1}=\log m_1\,,\qquad y_q=\log\frac{q}{1-q}\,,\qquad y_z=\log z\,.
\end{equation}
Spin magnitudes and tilts are poorly measured for most events, with their posteriors often extending to the boundaries of their domains ($\chi\to0,1$ or $\cos\theta\to\pm1$).
We found that logit or inverse hyperbolic tangent transformations map these regions to the tails of the mixture, increasing the number of components needed and potentially biasing population inference.
For this reason, similar to Ref.~\cite{Golomb:2021tll}, we instead transform each spin parameter using
\begin{equation}\label{eq: ecdf probit map}
    y_{p}=\Phi^{-1}\!\left[\hat{F}(\theta_p)\right]\,,
\end{equation}
where $\Phi^{-1}$ is the inverse CDF of the standard normal distribution and $\hat{F}$ is the weighted marginal empirical CDF of $\theta_p$.
This map transforms each one-dimensional spin marginal of the weighted samples into a standard normal distribution, allowing the mixture to be fitted with fewer components.
Finally, we standardize each coordinate as $\tilde{y}_p=(y_p-\bar{y}_p)/s_p$, where $\bar{y}_p$ and $s_p$ are the weighted mean and standard deviation of the samples in coordinate $y_p$.

When mapped back to the physical parameter space, the target density is given by
\begin{equation}\label{eq: gmm density}
    \frac{p(\theta|d_i)}{\ppe(\theta)} \simeq \hat{g}_i(\theta)=\left[\prod_{p=1}^{P}\frac{1}{s_p}\left|\frac{\dd{y_p}}{\dd{\theta_p}}\right|\right]\sum_{k=1}^{K}\eta_k\,\mathcal{N}(\tilde{y};\mu_k,\Sigma_k)\,,
\end{equation}
where $K$ is the number of GMM components,
$\mathcal{N}(\cdot\,;\mu_k,\Sigma_k)$ is the multivariate normal density with mean $\mu_k$ and covariance $\Sigma_k$, and
$\eta_k$ are the mixture weights, with $\sum_k\eta_k=1$.

We use $K=50$ components for all events to balance fit accuracy against the computational cost of evaluating the population likelihood. We verified that varying $K$ between 10 and 300 does not significantly impact our results; conversely, the
choice of the transformation in Eq.~\eqref{eq: ecdf probit map} proved to be crucial to avoid biases.
During inference, each mixture is evaluated in parallel on a GPU at the $N_{\lambda, i}$ population samples for the corresponding event, so both evaluation time and memory usage scale with $K$.
The fitting procedure is implemented in \textsc{\small PyTorch}~\cite{Paszke:2019xhz}.

An example of our GMM fit is illustrated in Fig.~\ref{fig: gmm fit} for GW150914, where we show
the agreement between the GMM (purple) and the prior-weighted PE samples (blue).

\subsection{Selection function}
\begin{figure*}[!ht]
    \centering
    \includegraphics[width=\textwidth]{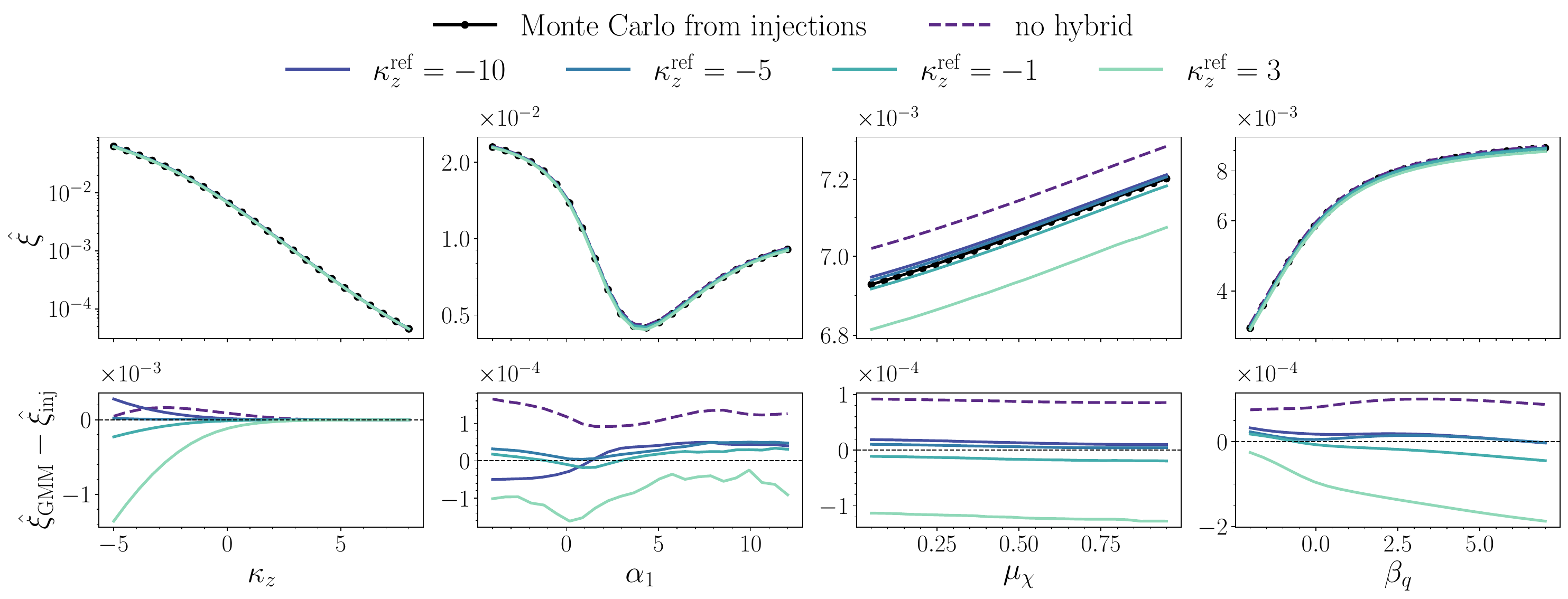}
    \caption{Comparison of selection function estimates from the injection-based estimator in Eq.~\eqref{eq: Monte Carlo Pdet estimator} (black points), the GMM without hybridization (blue dashed curves), and the GMM with redshift hybridization for different values of $\kappa_z^{\rm ref}$ (solid curves).
The estimates are shown as functions of selected parameters $\lambda$ of the GWTC-5.0 \defaultBBH population model~\cite{LIGOScientific:2026ctl}.
Each parameter is varied while the others are held fixed at their maximum-likelihood values from the analysis of Ref.~\cite{LIGOScientific:2026ctl}.}
    \label{fig:pdet_estimator_comparison}

\end{figure*}

We require an estimator of the selection function that can be evaluated at arbitrary points in parameter space, and not only at the fixed locations of the injections as in  Eq.~\eqref{eq: Monte Carlo Pdet estimator}.
To this end, a viable solution is to employ emulators for $p(\det|\theta)$ either based
on physically motivated analytical models~\cite{Lorenzo-Medina:2024opt, Essick:2023toz, Mould:2023eca, Gerosa:2024isl}, which make simplifying assumptions about the detection model,
or on neural networks~\cite{Gerosa:2020pgy, Callister:2024qyq, Gassert:2026ugv},
which typically require injections both above and below the detection threshold to be available.
The public data product released by the LVK Collaboration does not contain such a complete set, as a large number of ``hopeless'' sources are not injected into noise in order to save computational time~\cite{Essick:2025zed}. To avoid the computational cost of generating a new set of injections, we instead adopt a strategy similar to that introduced in Ref.~\cite{Talbot:2020oeu}, allowing estimates of the selection function by fitting a density estimator to the found injections alone.

The injections released by the LVK Collaboration for the GWTC-5.0 population analysis~\cite{LIGOScientific:2026ctl, ligo_scientific_collaboration_2026_19500052} are divided into $N=5$ sets, indexed by $r$, each drawn from its own reference distribution $q_r(\theta)$:
separate sets of semianalytic injections for the O1 and O2 observing runs~\cite{Essick:2023toz},
two sets targeting stellar-mass binary BHs and intermediate-mass BHs in O3~\cite{KAGRA:2021vkt, KAGRA:2021duu, ligo_scientific_collaboration_and_virgo_2021_5546676},
and a set for the O4 observing run~\cite{Essick:2025zed}.
We refer the reader to these works for definitions of the reference distributions $q_r(\theta)$ and to Ref.~\cite{2021RNAAS...5..220E} for the method
used to combine injections from different sets.
An O3 or O4 injection is considered found if its false alarm rate is below $1\,\mathrm{yr}^{-1}$ in at least one search, while a semianalytic O1 or O2 injection is considered found if its network signal-to-noise ratio $\rho$ exceeds $10$~\cite{Essick:2023toz,Mould:2023eca}.
The numbers of found injections are $N_r\simeq1.1\times10^{5}$, $2.6\times10^{5}$, $8.1\times10^{4}$, $1.5\times10^{4}$, and $9.9\times10^{5}$ for the O1, O2, O3$_{\rm BBH}$, O3$_{\rm IMBH}$, and O4 sets, respectively.

Separating the contributions of the different injection sets, the estimator of Eq.~\eqref{eq: Monte Carlo Pdet estimator} becomes
\begin{equation}\label{eq: pdet split by run}
    \hat{\xi}_{\rm inj}(\lambda) = \frac{1}{N_{\rm inj}}\sum_{r=1}^{N}\sum_{j=1}^{N_r} w_{rj}\,\frac{\ppop(\theta_{rj}|\lambda)}{q_r(\theta_{rj})}\,,
\end{equation}
where $\theta_{rj}$ are the found injections of set $r$, distributed as
\begin{equation}\label{eq: found injection density}
    q_r(\theta|\mathrm{det}) = \frac{p(\mathrm{det}|\theta)\,q_r(\theta)}{\int\dd{\theta'}\,p(\mathrm{det}|\theta')\,q_r(\theta')}\,.
\end{equation}
Writing the weight of each injection as a function $w_r(\theta)$ of its parameters, the inner sum of Eq.~\eqref{eq: pdet split by run} is a Monte Carlo estimate of
\begin{equation}\label{eq: pdet inner sum}
    \sum_{j=1}^{N_r} w_{rj}\,\frac{\ppop(\theta_{rj}|\lambda)}{q_r(\theta_{rj})} \simeq N_r\!\int\!\dd{\theta}\,w_r(\theta)\,q_r(\theta|\mathrm{det})\,\frac{\ppop(\theta|\lambda)}{q_r(\theta)}\,.
\end{equation}
A normalized density estimator fitted to the found injections with weights $w_{rj}$ then approximates
\begin{equation}\label{eq: weighted found density}
    \hat{q}_r(\theta|\mathrm{det}) \simeq \frac{w_r(\theta)\,q_r(\theta|\mathrm{det})}{\int\dd{\theta'}\,w_r(\theta')\,q_r(\theta'|\mathrm{det})}\,,
\end{equation}
where the denominator is in turn estimated by $N_r^{-1}\sum_{j}w_{rj}$.
Substituting Eq.~\eqref{eq: weighted found density} into Eq.~\eqref{eq: pdet inner sum}, we obtain
\begin{equation}\label{eq: pdet integral form}
    \hat{\xi}_{\rm inj}(\lambda) \simeq \frac{1}{N_{\rm inj}}\sum_{r=1}^{N}\left(\sum_{j=1}^{N_r}w_{rj}\right)\int\dd{\theta}\,\ppop(\theta|\lambda)\,\frac{\hat{q}_r(\theta|\mathrm{det})}{q_r(\theta)}\,.
\end{equation}
The ratio $\hat{q}_r(\theta|\mathrm{det})/q_r(\theta)$ is proportional to $w_r(\theta)\,p(\mathrm{det}|\theta)$:
the fit to the found injections therefore acts as an emulator of the detection probability,
which requires neither the missed injections nor an evaluation of $\ppop(\theta|\lambda)$.

The selection-function estimator therefore becomes
\begin{align}
    \hat{\xi}(\lambda) &= \frac{1}{N_\mathrm{inj}} \sum_{r=1}^{N}\left[
    \left( \sum_{j=1}^{N_r} w_{rj} \right)
    \frac{1}{N_\lambda} \sum_{k=1}^{N_\lambda} \frac{\hat{q}_r(\theta_k|\mathrm{det})}{q_r(\theta_k)}\right]\,,
    \notag\\
    \theta_{k} &\sim \ppop(\theta|\lambda)\,,
    \label{eq: selection effects estimator from pop samples}
\end{align}
where $\hat{q}_r(\theta|\mathrm{det})$ is the density estimator for the found injections drawn from the $r$-th distribution,
the weight $w_{rj}$ refers to the $j$-th found injection of the $r$-th set,
and the integral is estimated with $N_\lambda$ samples from $\ppop(\theta|\lambda)$.

As with the individual-event posteriors, we fit each $\hat{q}_r(\theta|\mathrm{det})$ separately using a GMM in unbounded coordinates.
Since the analytic forms of $q_r(\theta)$ are known~\cite{Essick:2023toz}, we can map the physical injection parameters $\theta$ to unbounded coordinates $y$ through
a bijective transformation $y=T_r(\theta)$, such that the transformed density is approximately standard normal.
Specifically, we construct $T_r$
using the analytical marginal CDF transforms of $q_r$, followed by the inverse standard normal CDF $\Phi^{-1}$ applied componentwise.
For the mass ratio only, we use instead the CDF conditioned on $m_1$, to account for the
$m_1$-dependent limit on allowed mass ratios.
By construction, $T_r$ maps $q_r$ onto a standard normal,
\begin{equation}
    q_r(\theta) = \phi\!\left(y\right)\left|\frac{\partial y}{\partial\theta}\right|\,,
    \qquad \phi(y)=\prod_p\phi(y_p)\,,
\end{equation}
where $\phi$ is the standard normal density, and $\phi(y)$ its product over the coordinates of $y$.
If $D_r(y)$ is the GMM density of the weighted found injections in these coordinates,
then $\hat{q}_r(\theta|\mathrm{det})=D_r(y)\,|\partial y/\partial\theta|$, and the Jacobians cancel in the ratio, so that
\begin{equation}\label{eq: pdet gmm ratio}
    \frac{\hat{q}_r(\theta|\mathrm{det})}{q_r(\theta)} = \frac{D_r(y)}{\phi(y)}\,.
\end{equation}
The GMM therefore only needs to model how selection effects distort a standard normal distribution.
Although $q_r(\theta_k)$ appears in Eq.~\eqref{eq: selection effects estimator from pop samples}, Eq.~\eqref{eq: pdet gmm ratio} evaluates the ratio at $y=T_r(\theta_k)$, so the injection distribution enters only through
the CDFs that define $T_r$; neither the density $q_r$ nor the Jacobian of $T_r$ needs to be computed.

We fit $D_r$ using the same weighted EM algorithm used to fit the individual-event posteriors (cf. Sec.~\ref{sec: gmm fit posteriors}).
To accurately reproduce the injection-based estimator of Eq.~\eqref{eq: Monte Carlo Pdet estimator},
we found the choice of the number of components to be crucial, and selected it by minimizing the Bayesian information criterion (BIC)~\cite{Schwarz:1978tpv},
giving $K=8$, 14, 6, 4, and 20 for the O1, O2, O3$_{\rm BBH}$, O3$_{\rm IMBH}$, and O4 sets,
respectively.

By substituting Eq.~\eqref{eq: pdet gmm ratio} into Eq.~\eqref{eq: selection effects estimator from pop samples}, we then
obtain the final form of our GMM-based estimator of the selection function:
\begin{align}
    \hat{\xi}_{\rm GMM}(\lambda) &= \frac{1}{N_{\rm inj}}\sum_{r=1}^{N}\left(\sum_{j=1}^{N_r}w_{rj}\right)
    \,\frac{1}{N_\lambda}\sum_{k=1}^{N_\lambda}\frac{D_r(y_{rk})}{\phi(y_{rk})}\,,
    \notag\\
    \theta_{k} &\sim \ppop(\theta|\lambda)\,,
    \notag\\
    y_{rk}&=T_r(\theta_k)\,.
    \label{eq: pdet gmm estimator}
\end{align}
The samples $\theta_k$ are drawn once per likelihood call from the population model and used for all injection sets.

\subsection{Hybrid redshift sampling}\label{sec: hybrid redshift sampling}

The comoving volume and merger rate typically increase with $z$, so most samples drawn from realistic populations lie at high redshift,
where the detection probability is negligible. Most found injections, however, are at low redshift.
The estimate in Eq.~\eqref{eq: pdet gmm estimator} is therefore often dominated by a few low-redshift samples,
resulting in a large variance.
In the very specific cases in which
\begin{enumerate}
    \item[(a)] the population model factorizes as $\ppop(\theta|\lambda)=\ppop(z|\lambda)\,\ppop(\theta_{\rm rest}|\lambda)$, where $\theta_{\rm rest}$ denotes the parameters other than redshift, and
    \item[(b)] the redshift distribution $\ppop(z|\lambda)$ is tractable, as is the case for a merger rate that evolves as a power law in $1+z$~\cite{Fishbach:2018edt},
\end{enumerate}
it is however possible to increase the effective sample size, by augmenting the estimator
with the same hybrid scheme introduced in Ref.~\cite{Callister:2024qyq}.
In this scheme, redshift samples are drawn from a fixed reference distribution $p_{\rm ref}(z)$ and reweighted by $\ppop(z|\lambda)/p_{\rm ref}(z)$.

We draw $N_\lambda=10^{6}$ redshift samples $z_k$ from the reference distribution
\begin{align}
p_{\rm ref}(z)\propto \frac{\dd{V_c}}{\dd{z}} (1+z)^{\kappa_z^{\rm ref} -1}\,,
\end{align}
where $V_c$ is the comoving volume. We truncate this distribution at $z=3$~\cite{Essick:2025zed}. 
This distribution can be tuned to place more samples at low redshift than typical merger-rate models.
The remaining parameters are drawn from the population model, and the resulting estimator is
\begin{align}
    \hat{\xi}_{\rm GMM}(\lambda) &= \frac{1}{N_{\rm inj}}\sum_{r=1}^{N}\left(\sum_{j=1}^{N_r}w_{rj}\right)\notag \\
    &\times\frac{1}{N_\lambda}\sum_{k=1}^{N_\lambda}\frac{D_r(y_{rk})}{\phi(y_{rk})}\,\frac{\ppop(z_k|\lambda)}{p_{\rm ref}(z_k)}\,,
    \label{eq: pdet hybrid estimator}
\end{align}
which differs from Eq.~\eqref{eq: pdet gmm estimator} only by the reweighting factor $\ppop(z_k|\lambda)/p_{\rm ref}(z_k)$.

In Fig.~\ref{fig:pdet_estimator_comparison}, we compare estimates of the selection function obtained using the standard approach of Eq.~\eqref{eq: Monte Carlo Pdet estimator} and the GMM-based estimator of Eq.~\eqref{eq: pdet hybrid estimator}.
For this comparison, we adopt the GWTC-5.0 \defaultBBH population model (see Sec.~\ref{sec: comparison} and Ref.~\cite{LIGOScientific:2026ctl} for details).
We vary each model parameter $\lambda$ in turn, keeping the others fixed at their maximum-likelihood values from the analysis of Ref.~\cite{LIGOScientific:2026ctl}.
We show the dependence of the selection function on $\lambda$ for different values of $\kappa_z^{\rm ref}$.
The hybrid scheme typically improves agreement with the injection-based estimator. Based on these residuals, we adopt $\kappa_z^{\rm ref}=-1$ for the remainder of this work.

Specifically, in the applications provided below, this redshift hybridization is used in Secs.~\ref{sec: comparison} and~\ref{sec: hierarchical mergers}, but not in Sec.~\ref{sec: pop synth},
because populations from realistic astrophysical simulations do not satisfy the conditions above.

\subsection{Monte Carlo variance and restricted sampling}\label{sec: variance and restricted sampling}
Monte Carlo estimates have an intrinsic uncertainty due to the finite number of random draws. 
The estimated log-likelihood is a sum of logarithms of Monte Carlo integrals [cf. Eqs.~\eqref{eq: standard monte carlo likelihood estimator} and~\eqref{eq: sampling p_pop monte carlo likelihood estimator}].
Propagating the errors of these integrals to first order gives~\cite{Essick:2022ojx}
\begin{align}
    \mathrm{Var}[\log\hat{\mathcal{L}}] &\simeq \sum_{i,j=1}^{\Nobs}\frac{\mathrm{Cov}[\hat{\mathcal{L}}_i,\hat{\mathcal{L}}_j]}{\hat{\mathcal{L}}_i\,\hat{\mathcal{L}}_j} + \Nobs^2\,\frac{\mathrm{Var}[\hat{\xi}]}{\hat{\xi}^2}\notag \\
    &-2\Nobs\sum_{i=1}^{\Nobs}\frac{\mathrm{Cov}[\hat{\mathcal{L}}_i,\hat{\xi}]}{\hat{\mathcal{L}}_i\,\hat{\xi}}\,,
    \label{eq: log-likelihood variance full}
\end{align}
where the first term contains cross terms between different events when $i\neq j$ and the last one contains cross terms between events and selection effects.

Both types of cross terms in Eq.~\eqref{eq: log-likelihood variance full} vanish when the integrals involved are computed from
independent sets of samples.
This is the case in the standard approach, where each event term is computed from the PE samples of that specific event and the selection function is computed using injections [cf.~Eqs.~\eqref{eq: standard monte carlo likelihood estimator} and~\eqref{eq: Monte Carlo Pdet estimator}]. In this case, Eq.~\eqref{eq: log-likelihood variance full} reduces to
\begin{equation}\label{eq: log-likelihood variance}
    \mathrm{Var}[\log\hat{\mathcal{L}}] \simeq \sum_{i=1}^{\Nobs}\frac{\mathrm{Var}[\hat{\mathcal{L}}_i]}{\hat{\mathcal{L}}_i^2} + \Nobs^2\,\frac{\mathrm{Var}[\hat{\xi}]}{\hat{\xi}^2}\,,
\end{equation}
which is the expression most commonly adopted in the literature (cf. Refs.~\cite{Talbot:2023pex, Mancarella:2025uat, Heinzel:2025ogf, LIGOScientific:2026ctl}).
In our case, however, this approximation does not hold in general. Specifically, population samples can be drawn independently for each event and for the selection function in the simpler explorations of Secs.~\ref{sec: comparison} and~\ref{sec: hierarchical mergers}, but not in the more realistic population-synthesis applications of Sec.~\ref{sec: pop synth}, where the sampled binaries are provided by a fixed set of simulated sources.

Controlling this variance is crucial to avoid biasing the posterior of the population parameters and to ensure convergence of the posterior sampling algorithm.
A straightforward solution is to increase the number of samples used to compute the Monte Carlo averages, but this can be computationally expensive and is not always feasible.
For instance, when estimating the selection function using Eq.~\eqref{eq: Monte Carlo Pdet estimator}, the effective sample size is limited by the total number of injections $N_{\rm inj}$.
In practice, a common strategy is to impose an upper threshold on $\mathrm{Var}[\log\hat{\mathcal{L}}]$ during sampling, preventing the sampler from exploring regions of hyperparameter space with high variance.
Applying this threshold effectively modifies the likelihood, 
thus violating the assumptions of the inhomogeneous Poisson process used to formulate the population inference problem.

Addressing the variance becomes even more important in the case of the estimator of Eq.~\eqref{eq: sampling p_pop monte carlo likelihood estimator}. As pointed out by Refs.~\cite{Mould:2023ift, Mancarella:2025uat}, posterior distributions of individual events typically have a much narrower support than the entire population model, especially in the mass parameters.

Evaluating the integral as in Eq.~\eqref{eq: sampling p_pop monte carlo likelihood estimator} by drawing samples from the population model $\ppop(\theta|\lambda)$ might therefore result in a large fraction of samples that do not contribute to the individual-event likelihoods, hence increasing the variance and lowering the sampling efficiency.

The selection term contribution to the variance can be reduced by the hybrid redshift sampling of Sec.~\ref{sec: hybrid redshift sampling} when applicable, or by increasing the number of population samples.
We instead reduce the dominant contribution from single-event likelihoods
by using the PE posteriors to guide sampling from the population model. Specifically, we restrict the integral in Eq.~\eqref{eq: standard hierarchical likelihood} to a region of parameter space where the single-event posteriors (and thus likelihoods) have support.

Let $R_i$ be a region of parameter space $\theta$. We define
\begin{equation}\label{eq: restricted population}
    \ppop(\theta|\lambda, R_i) = \frac{\ppop(\theta|\lambda)\,\Theta(\theta\in R_i)}{P(R_i|\lambda)}\,,
\end{equation}
where $P(R_i|\lambda)=\int_{R_i}\dd{\theta}\,\ppop(\theta|\lambda)$ is the population probability mass within $R_i$.
If $p(\theta|d_i)$ is negligible outside $R_i$, the integral over $\theta$ in Eq.~\eqref{eq: standard hierarchical likelihood} can be restricted to $R_i$.
Multiplying and dividing the integrand by $P(R_i|\lambda)$ then yields the final estimator
\begin{align}
\hat{\mathcal{L}}_R(\lambda)
& =
\hat{\xi}_{\rm GMM}(\lambda)^{-\Nobs}
\prod_{i=1}^{\Nobs}
\frac{P(R_i|\lambda)}{N_{\lambda, i}}
\sum_{k=1}^{N_{\lambda, i}}
\frac{p(\theta_{ki}|d_i)}{\ppe(\theta_{ki})}
\,, \notag\\
\theta_{ki} & \sim \ppop(\theta|\lambda,R_i)
\,, \label{eq: restricted population likelihood estimator}
\end{align}
with $\hat{\xi}_{\rm GMM}(\lambda)$ the selection function estimator of either Eq.~\eqref{eq: pdet gmm estimator} or~\eqref{eq: pdet hybrid estimator}.
The restriction to $R_i$ reduces the relative variance of each event term by at least a factor $P(R_i|\lambda)$.
To show this, let $\mathcal{L}_i(\lambda)$ be the integral on the right-hand side of Eq.~\eqref{eq: standard hierarchical likelihood event},
whose Monte Carlo estimator $\hat{\mathcal{L}}_i(\lambda)$ is the term in square brackets in Eq.~\eqref{eq: sampling p_pop monte carlo likelihood estimator}.
Splitting this integral over $R_i$ and its complement $\bar R_i$ gives
\begin{equation}\label{eq: likelihood split}
    \mathcal{L}_i(\lambda) = \mathcal{L}_{R_i}(\lambda) + \mathcal{L}_{\bar R_i}(\lambda)\,.
\end{equation}
In Eq.~\eqref{eq: restricted population likelihood estimator}, the term of event $i$ is the Monte Carlo estimator of $\mathcal{L}_{R_i}$,
\begin{equation}\label{eq: restricted event estimator}
    \hat{\mathcal{L}}_{R_i} = P(R_i|\lambda)\,\tilde{\mathcal{L}}_{R_i} \,,
\end{equation}
where $\tilde{\mathcal{L}}_{R_i}$ is the average of $p(\theta|d_i)/\ppe(\theta)$ over the $N_{\lambda, i}$ samples $\theta_{ki}\sim\ppop(\theta|\lambda,R_i)$, i.e., without the factor $P(R_i|\lambda)$.
We denote by $\tilde{\mathcal{L}}_{\bar R_i}$ the analogous average over $N_{\lambda, i}$ samples drawn from $\ppop(\theta|\lambda,\bar R_i)$.

The advantage of Eq.~\eqref{eq: restricted event estimator} is that all $N_{\lambda, i}$ samples contribute to the average.
When sampling from $\ppop(\theta|\lambda)$, as in Eq.~\eqref{eq: sampling p_pop monte carlo likelihood estimator}, each sample instead falls in $R_i$ only with probability $P(R_i|\lambda)$.
Denoting by $N_{R_i}\leq N_{\lambda, i}$ the number of samples that fall in $R_i$, and since $p(\theta|d_i)$ is negligible in $\bar R_i$, the unrestricted estimator can be written as
\begin{equation}\label{eq: unrestricted product}
    \hat{\mathcal{L}}_i \simeq \frac{N_{R_i}}{N_{\lambda, i}}\,\frac{1}{N_{R_i}}\sum_{\theta_{ki}\in R_i}\frac{p(\theta_{ki}|d_i)}{\ppe(\theta_{ki})}\,.
\end{equation}
The samples falling in $R_i$ are distributed as $\ppop(\theta|\lambda,R_i)$, so the second factor has the same form as $\tilde{\mathcal{L}}_{R_i}$,
but it averages over only $N_{R_i}\simeq N_{\lambda, i} P(R_i|\lambda)$ samples instead of $N_{\lambda, i}$.

To quantify the resulting gain, we apply the law of total variance to Eq.~\eqref{eq: sampling p_pop monte carlo likelihood estimator}, conditioning on whether each sample falls in $R_i$. This gives
\begin{align}
    \mathrm{Var}&[\hat{\mathcal{L}}_i] = P(R_i|\lambda)\,\mathrm{Var}[\tilde{\mathcal{L}}_{R_i}] + \left[1-P(R_i|\lambda)\right]\mathrm{Var}[\tilde{\mathcal{L}}_{\bar R_i}]
   \notag \\
   & + \frac{P(R_i|\lambda)\left[1-P(R_i|\lambda)\right]}{N_{\lambda, i}}
    \left[\frac{\mathcal{L}_{R_i}}{P(R_i|\lambda)}-\frac{\mathcal{L}_{\bar R_i}}{1-P(R_i|\lambda)}\right]^2
     \notag \\
    &\simeq P(R_i|\lambda)\,\mathrm{Var}[\tilde{\mathcal{L}}_{R_i}] +\frac{1-P(R_i|\lambda)}{N_{\lambda, i}\,P(R_i|\lambda)}\,\mathcal{L}_i^2\,,
    \label{eq: variance split}
\end{align}
where we used Eq.~\eqref{eq: likelihood split} and neglected
both $\mathrm{Var}[\tilde{\mathcal{L}}_{\bar R_i}]$ and $\mathcal{L}_{\bar R_i}$ because $p(\theta|d_i)$ is negligible in $\bar R_i$ by construction, so that $\mathcal{L}_{R_i}\simeq\mathcal{L}_i$.
The second term in the latest line of Eq.(\ref{eq: variance split}) equals $\mathcal{L}_i^2\,\mathrm{Var}[N_{R_i}/N_{\lambda, i}]/P(R_i|\lambda)^2$, i.e., it arises from the variance of the prefactor
$N_{R_i}/N_{\lambda, i}$ of Eq.~\eqref{eq: unrestricted product}, which estimates $P(R_i|\lambda)$ by counting over $N_{\lambda, i}$ draws.
The restricted estimator of Eq.~\eqref{eq: restricted event estimator} has the same structure, with a factor $\hat{P}(R_i|\lambda)$ that can either be computed from the CDFs of the population model
or estimated by counting. Its variance is therefore
    \begin{align}
    \mathrm{Var}[\hat{\mathcal{L}}_{R_i}] &\simeq P(R_i|\lambda)^2\,\mathrm{Var}[\tilde{\mathcal{L}}_{R_i}] + \frac{\mathrm{Var}[\hat{P}(R_i|\lambda)]}{P(R_i|\lambda)^2}\,\mathcal{L}_i^2 \notag \\
    &\leq P(R_i|\lambda)\,\mathrm{Var}[\hat{\mathcal{L}}_i]\leq \mathrm{Var}[\hat{\mathcal{L}}_i]\,,
    \label{eq: variance restricted}
\end{align}
where the first inequality follows from comparing with the last line of Eq.~\eqref{eq: variance split} term by term.
The first term is a factor $P(R_i|\lambda)$ smaller, because for the restricted estimator all $N_{\lambda, i}$ samples contribute to the average,
rather than only the $\simeq N_{\lambda, i}\,P(R_i|\lambda)$ that fall in $R_i$ as in Eq.~\eqref{eq: unrestricted product}.
In the second term, $\mathrm{Var}[\hat{P}(R_i|\lambda)]$ vanishes if $P(R_i|\lambda)$ is computed exactly,
while counting requires at least $N_{\lambda, i}/P(R_i|\lambda)$ draws to place $N_{\lambda, i}$ samples in $R_i$, so that $\mathrm{Var}[\hat{P}(R_i|\lambda)]\leq P(R_i|\lambda)\,\mathrm{Var}[N_{R_i}/N_{\lambda, i}]$.
For the same number $N_{\lambda, i}$ of evaluations of $p(\theta|d_i)/\ppe(\theta)$, which dominate the computational cost,
the restricted estimator therefore has a variance smaller by at least a factor $P(R_i|\lambda)$.

The restriction therefore reduces the variance by at least a factor $P(R_i|\lambda)$ for the same number $N_{\lambda, i}$ of evaluations of $p(\theta|d_i)/\ppe(\theta)$,
which dominate the computational cost.

This gain in the variance comes at the price of a truncation bias. From Eq.~\eqref{eq: likelihood split}, the expectation of $\hat{\mathcal{L}}_{R_i}$ is
\begin{equation}\label{eq: restricted mean}
    \mathcal{L}_{R_i} = (1-\epsilon_i)\,\mathcal{L}_i\,,\qquad \epsilon_i=\frac{\mathcal{L}_{\bar R_i}}{\mathcal{L}_i}\,,
\end{equation}
with $\epsilon_i$ being the fraction of $\mathcal{L}_i$ contributed by $\bar R_i$.
The log-likelihood is therefore shifted by $\sum_i\log(1-\epsilon_i)\simeq-\sum_i\epsilon_i$.
This highlights the importance of choosing $R_i$ carefully enough to enclose both the bulk and the tail of the posterior support,
so that this sum varies by much less than unity across the hyperparameter posterior $p(\lambda|\mathcal{D})$.

\begin{figure}[!t]
    \centering
    \includegraphics[width=\columnwidth]{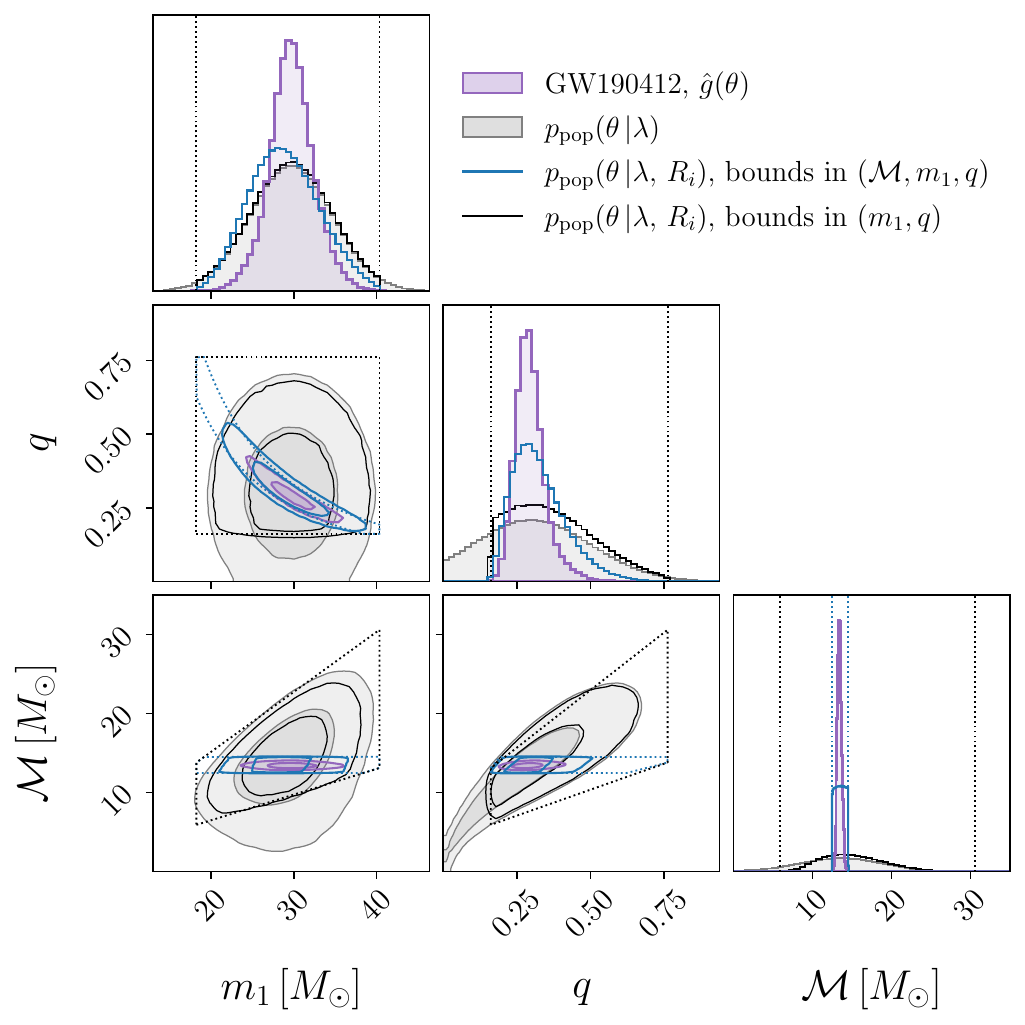}
    \caption{Illustrative example of our restricted sampling scheme for GW190412 and a Gaussian toy population model in $(m_1,q)$.
Most samples drawn directly from the population model (gray) fall outside the posterior support (purple). Contours enclose the 50\% and 90\% credible regions. Restricting sampling to the $(m_1,q)$ box (black dashed) improves the overlap of the sampled distribution (black solid) with the posterior, but still produces samples outside its correlated support.
The best agreement is obtained by imposing simultaneous constraints on $\mathcal{M}$, $m_1$, and $q$ (blue dashed). The resulting sampled distribution (blue solid) more closely follows the correlated posterior support.}
    \label{fig: restricted sampling}
\end{figure}

We therefore apply this restriction strategy only to the mass and redshift parameters ($m_1$, $q$, and $z$), since the spin posteriors span their full allowed ranges for most events.
We construct an initial region $R_i$ using the $99.9\%$ credible intervals of the one-dimensional marginals of $p(\theta|d_i)/\ppe(\theta)$, computed from samples drawn from the GMM fits $\hat{g}_i(\theta)$ described in Sec.~\ref{sec: gmm fit posteriors}.
We refine this region by accounting for the strong correlations between the mass parameters.
For most events, the posterior support is better described in terms of the chirp mass $\mathcal{M}=(m_1m_2)^{3/5}(m_1+m_2)^{-1/5}$ and the mass ratio $q$ than in terms of $m_1$ and $q$.
A rectangular region in $(\mathcal{M},q)$, transformed back to $(m_1,q)$, therefore provides a more accurate approximation to the posterior support.
We combine the two approaches by defining $R_i$ as the region where $\mathcal{M}$, $m_1$, $q$, and $z$ each lie within the $99.9\%$ credible interval of their respective one-dimensional marginals.

The methods for drawing samples from $\ppop(\theta|\lambda,R_i)$ and obtaining $P(R_i|\lambda)$ depend on how the population model is specified.
For instance, in the case of simple models that can be specified in terms of cumulative distribution functions (CDFs),
samples can be drawn directly inside $R_i$ by inverse-transform sampling; $P(R_i|\lambda)$ then follows directly
from integrating the population over $R_i$.
In the more interesting case of models available only through samples, the restricted population is the subset of samples falling within $R_i$, and $P(R_i|\lambda)$
is given by their fraction.
For a population model that factorizes into independent blocks of parameters, the two approaches can be combined,
using CDFs for some blocks and samples for the others.
By construction, $P(R_i|\lambda) = \prod_p P(R_i|\lambda_p)$ is then a product of the probabilities of each block.

Provided that the model is sufficiently fast to sample, the restriction to $R_i$ allows us to increase the effective number of samples contributing to the Monte Carlo average,
hence reducing the variance of the estimator at the same likelihood cost.
We illustrate the effectiveness of this restriction in Fig.~\ref{fig: restricted sampling}, using a Gaussian population model

\begin{equation}
    \begin{split}
    \ppop(m_1, q|\mu_m, \sigma_m,\mu_q, \sigma_q) &= \mathcal{N}(m_1 | \mu_m, \sigma_m)\,\\
    &\times\mathcal{N}_{[0,1]}(q | \mu_q, \sigma_q)\,,
    \end{split}
\end{equation}with $\mu_m=30\,M_\odot$, $\sigma_m=5\,M_\odot$, $\mu_q=0.3$, and $\sigma_q=0.2$.
We consider the event GW190412, whose chirp mass is well measured.
Sampling within the full region $R_i$ provides better coverage of the correlated posterior support than either restricting samples to a box in $(m_1,q)$ alone or sampling directly from the population model.
Both alternatives are more likely to produce samples outside the posterior support, thus resulting in a larger Monte Carlo variance when inserted in the population likelihood.

\subsection{Summary}\label{sec: summary}
We summarize here the necessary steps to perform a population inference by sampling, rather than evaluating, the population model $\ppop(\theta|\lambda)$:

\begin{enumerate}
    \item Draw $\lambda$ from the hyperparameter prior $\pi(\lambda)$.
    \item For each event $i=1,\dots,\Nobs$, draw $N_{\lambda, i}$ samples from the population $\ppop(\theta|\lambda, R_i)$,
    within the restricted support, and compute $P(R_i|\lambda)$.
    \item Draw $N_\lambda$ samples from the full population $\ppop(\theta|\lambda)$
    and evaluate $\hat{\xi}_{\rm GMM}(\lambda)$ with Eq.~\eqref{eq: pdet gmm estimator}.
   \item  If using the hybrid redshift scheme, draw $z$ from a fixed reference distribution $p_{\rm ref}(z)$ and
   reweight them by $\ppop(z|\lambda)/p_{\rm ref}(z)$, as in Eq.~\eqref{eq: pdet hybrid estimator}.
    \item Compute the likelihood estimator $\hat{\mathcal{L}}_R(\lambda)$ of Eq.~\eqref{eq: restricted population likelihood estimator}.
\end{enumerate}

We now apply our population-sampling approach to three different population models on the latest public catalog of events,
GWTC-5.0~\cite{LIGOScientific:2026wfs}.

\begin{figure*}
    \centering
    \includegraphics[width=\textwidth]{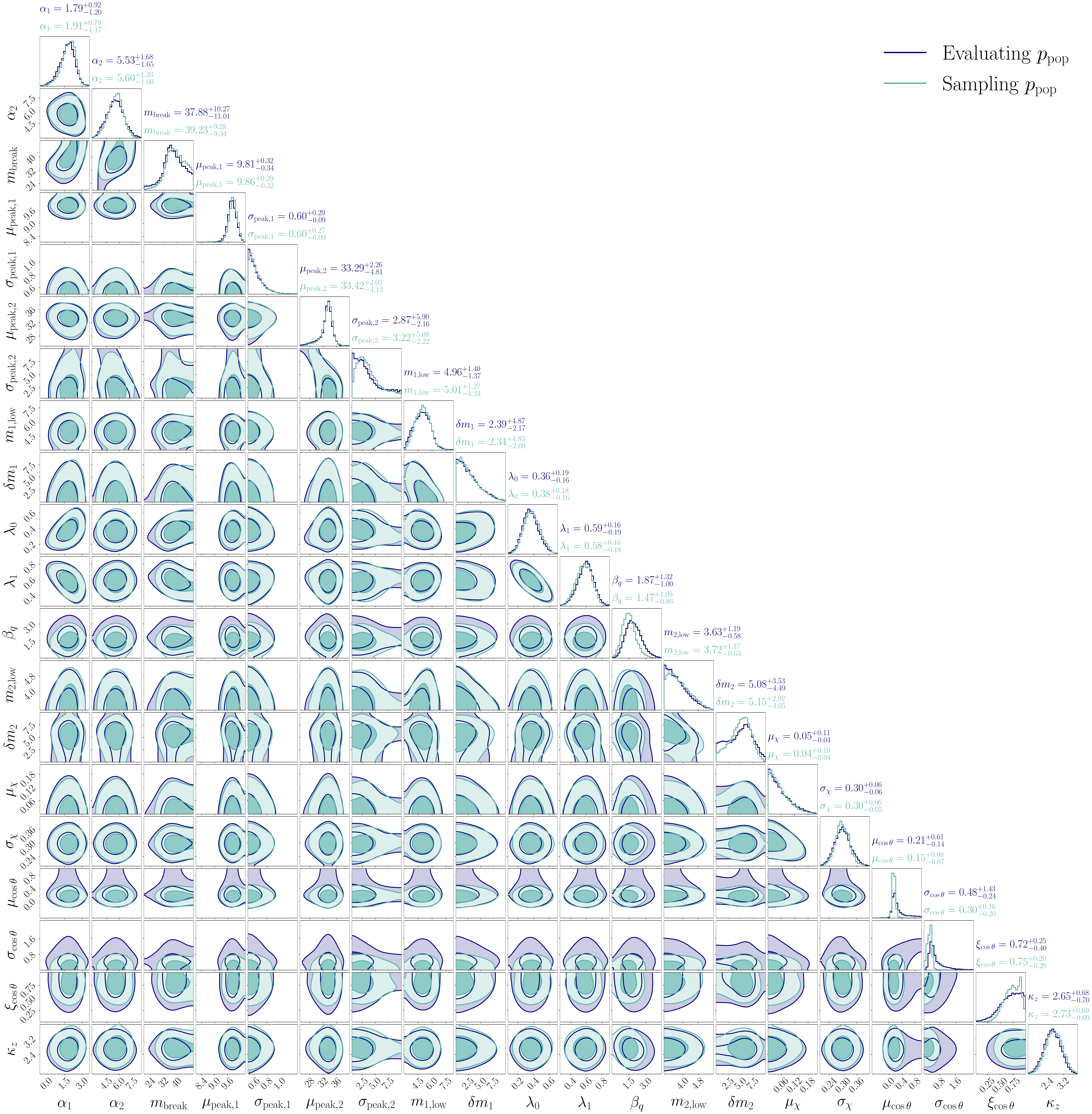}
    \caption{Posterior distribution of the population parameters of the \defaultBBH model inferred from GWTC-5.0. Blue shows the results of the standard inference, while teal shows those obtained by sampling the population model.
This 20-dimensional parametric model describes the distributions of masses, spin magnitudes, and spin directions. The axis labels denote the population parameters; see Refs.~\cite{LIGOScientific:2025pvj,LIGOScientific:2026ctl} for their definitions. Contours enclose the 50\% and 90\% credible regions; labels on the diagonal plots report medians and 90\% credible intervals.
The two methods yield largely consistent results, with some differences in the tails of the spin hyperparameter distributions, which are particularly sensitive to the variance of the likelihood estimator.
    }
    \label{fig: gwtc-5 default inference corner}
\end{figure*}

\section{Validation on a parametric model}\label{sec: comparison}

\subsection{Inference setup}

We first validate our method using the \defaultBBH model, the parametric population model adopted by the LVK Collaboration for the binary BH population analyses in both GWTC-4.0~\cite{LIGOScientific:2025pvj} and GWTC-5.0~\cite{LIGOScientific:2026ctl}.
This model assumes that the primary mass $m_1$ follows a mixture of a broken power law and two Gaussian components, while the mass ratio $q=m_2/m_1\leq 1$ follows a power law.
The spin magnitudes $\chi_{1,2}$ are drawn independently from the same truncated Gaussian distribution, while the cosines of the spin tilts $\cos\theta_{1,2}$ follow a mixture of an isotropic component and a truncated Gaussian component.
Finally, the merger rate density per unit comoving volume and source-frame time $\mathcal{R}(z)$ evolves as a power law in $1+z$.

We run the inference twice as outlined in Sec.~\ref{sec: strategies}, first with the standard Monte Carlo estimators that require evaluating $\ppop(\theta|\lambda)$, and second with the method developed here that requires sampling $\ppop(\theta|\lambda)$.
In the former case, we adopt the likelihood of Eq.~\eqref{eq: standard monte carlo likelihood estimator}
with selection effects estimated through Eq.~\eqref{eq: Monte Carlo Pdet estimator}, while in the latter we use Eq.~\eqref{eq: restricted population likelihood estimator} with the prescription
described in Sec.~\ref{sec: summary}.

We run the inferences with the \textsc{\small Bilby} package~\cite{Ashton:2018jfp}
using the \textsc{\small pocoMC} sampler~\cite{Karamanis:2022ksp},
an efficient sequential Markov chain Monte Carlo sampler, which
requires fewer likelihood evaluations than either standard Markov chain Monte Carlo or nested sampling, hence resulting in
a faster convergence to the posterior distribution. We validated this sampler by comparing the results against those obtained with \textsc{\small Dynesty}~\cite{Speagle:2019ivv}. The results shown hereafter are obtained by running \textsc{\small pocoMC}
with $\texttt{n}_\texttt{eff}=512$ particles and imposing a variance cut $\mathrm{Var}[\log \mathcal{L}]<4$ on the
likelihood estimator~\cite{LIGOScientific:2026ctl, Wolfe:2025coy}. The priors used follow those of Ref.~\cite{LIGOScientific:2026ctl}, except for the
redshift evolution parameter $\kappa_z$, for which we adopt a uniform prior in $\mathcal{U}(-5, 8)$ instead of $\mathcal{U}(-10, 10)$ for faster convergence.
For this analysis we draw $N_{\lambda, i}=2\times10^5$ population samples within each restricted region
$R_i$ and $N_\lambda=10^6$ samples from the full population to estimate the selection function with the hybrid scheme of Sec.~\ref{sec: hybrid redshift sampling}.

\subsection{Comparison}

\begin{figure}
    \centering
    \includegraphics[width=0.9\columnwidth]{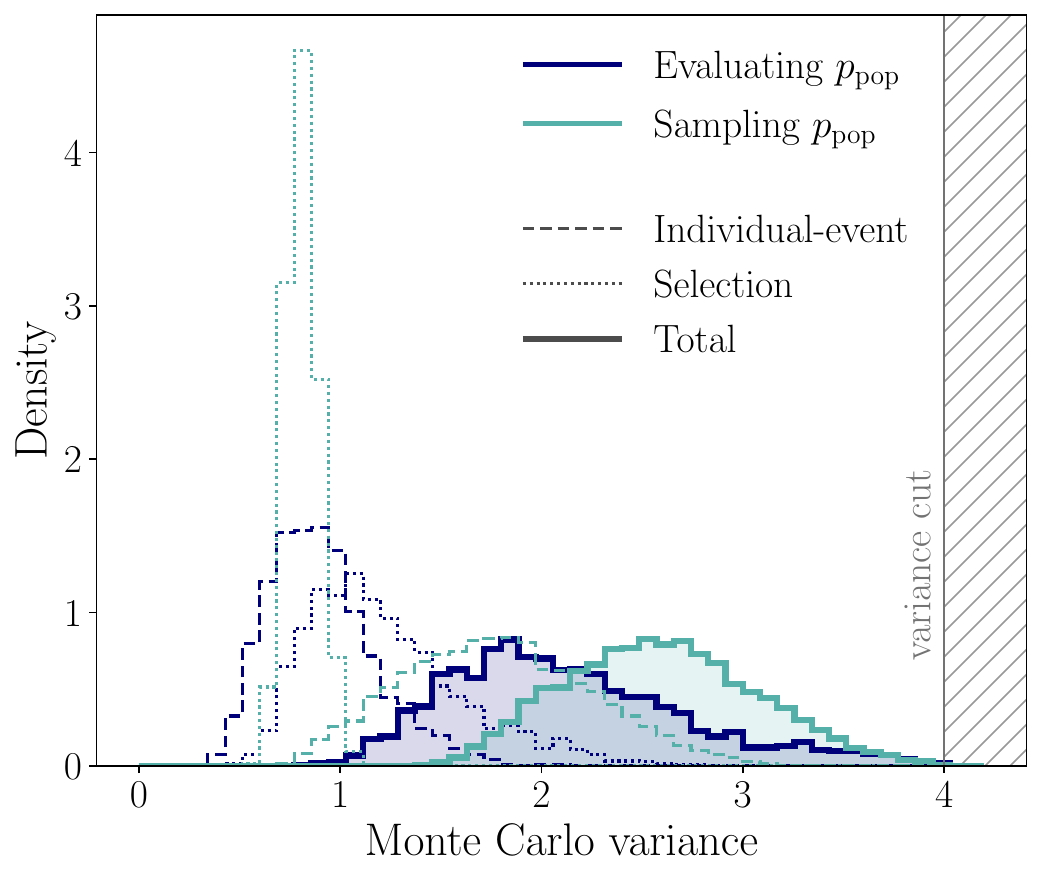}
    \caption{Distributions of the log-likelihood variance for the estimators in Eq.~\eqref{eq: standard monte carlo likelihood estimator} (blue) and Eq.~\eqref{eq: restricted population likelihood estimator} (teal), using the \defaultBBH model and GWTC-5.0 data.
Dashed and dotted histograms show the individual-event and selection contributions, respectively, while solid-line histograms show the total variance.
The higher total variance of our estimator is driven mainly by the individual-event contribution. Its selection contribution has a narrower distribution that peaks at lower values than that of the standard estimator.
The hatched region on the right is excluded during sampling by setting the likelihood to zero.
    }
    \label{fig: gwtc-5 default inference variance comparison}
\end{figure}

The posteriors from both inference runs are shown in Fig.~\ref{fig: gwtc-5 default inference corner}.
The posterior obtained by sampling $\ppop$ with our method largely overlaps with that obtained using the standard approach.
The largest
differences are visible in the spin-tilt parameters $(\mu_{\cos\theta}, \sigma_{\cos\theta})$: our posterior peaks at the dominant mode but does not extend as far into the tail.
Among the population parameters, those describing spins are notoriously sensitive to
Monte Carlo variance in the likelihood estimator~\cite{Mancarella:2025uat,LIGOScientific:2026ctl} and model assumptions~\cite{Vitale:2022dpa, Wolfe:2026meb, Wolfe:2026pjn}.
As illustrated in Fig.~\ref{fig: gwtc-5 default inference variance comparison}, our estimator has a slightly larger log-likelihood variance,
$\mathrm{Var}[\log \hat{\mathcal{L}}_R]=2.6^{+0.8}_{-0.7}$, than the standard estimator, $\mathrm{Var}[\log \hat{\mathcal{L}}]=2.0^{+1.3}_{-0.7}$; we quote medians and 90\% credible intervals.
This difference is driven mainly by the individual-event terms.
Notably, for the selection effects, our GMM estimator yields instead a narrower distribution of variances, concentrated at lower values, than the standard injection-based estimator
$\hat{\xi}_{\rm inj}$. In the standard approach, this is typically the other way around, with
the selection contribution larger than the one from individual events~\cite{Talbot:2023pex}.

The inferred spin hyperparameters also depend on the accuracy of the individual-event posterior fits.
Spin magnitudes and tilts are generally poorly measured, so their posteriors often extend to the boundaries of the parameter domain, $\chi_i\rightarrow 0,1$ and $\cos\theta_i\rightarrow\pm1$.
The accuracy of the fits near these boundaries therefore directly affects the likelihood.
Obtaining agreement between the posteriors for the two spin-tilt hyperparameters motivated the choices made in the GMM fitting procedure described in Sec.~\ref{sec: gmm fit posteriors}.
Overall, however, the agreement with the standard inference supports the reliability and accuracy of our method.

\section{Hierarchical mergers}\label{sec: hierarchical mergers}

We now turn to a population model that can be sampled but whose density cannot be evaluated directly.
Specifically, we consider hierarchical mergers, in which at least one binary component is the remnant of a previous merger (see Ref.~\cite{Gerosa:2021mno} for a review).
Such mergers are expected in dense stellar or gaseous environments, where merger remnants can pair with other BHs and merge again.
A key signature of hierarchical mergers is the remnant BH spin, which is typically larger than those of its progenitors. For sufficiently generic distributions of progenitor properties, the remnant spin distribution peaks at $\chi\approx0.7$~\cite{Fishbach:2017dwv, Gerosa:2017kvu}.
Hierarchical mergers can also populate the mass range above the maximum BH mass allowed by stellar evolution, which is expected to lie near $m\sim50\,M_\odot$ owing to pair-instability supernovae (PISN)~\cite{Belczynski:2016jno,Woosley:2016hmi,Farmer:2019jed}.

A consistent model of hierarchical mergers requires mapping the properties of BH progenitors to those of their remnants.
This mapping is known only through empirical fits to numerical-relativity (NR) simulations~\cite{Campanelli:2007ew,Barausse:2009uz,Barausse:2012qz,Gerosa:2016sys,Varma:2018aht} and allows us to generate samples from the hierarchical-merger subpopulation using samples from the first-generation subpopulation.
Transforming the corresponding probability density functions is much more difficult and generally cannot be done in closed form (but see Refs.~\cite{Doctor:2019ruh,Baxter:2021swn} for work in this direction).
Hierarchical mergers therefore provide a natural example of an ``easy to sample'' but ``hard to evaluate'' population model to fit to GW data.

\subsection{Mixture of black-hole generations}
\label{sec: hierarchical mergers population model}

We refer to BHs formed from stars as ``first-generation'' (1G) and to the remnants of mergers between 1G BHs as ``second-generation'' (2G).
We neglect contributions from higher generations (see, e.g., Ref.~\cite{Gerosa:2019zmo} for a discussion of the validity and limitations of this approximation).
We express the full population as a mixture of three branches, labeled by the generations of the two binary components,
\begin{equation}\label{eq: hierarchical mixture}
    \ppop(\theta|\lambda) = \sum_{b\in\{11,\,12,\,22\}} \eta_b\, p_b(\theta'|\lambda)\,p_b(z|\lambda)\,,
\end{equation}
where $\theta'$ denotes the parameters of Eq.~\eqref{eq: posterior parameters} other than $z$.
The labels $b=11$, $12$, and $22$ denote 1G+1G, 1G+2G, and 2G+2G binaries, respectively.
The branching ratios satisfy $\eta_b\geq0$ and $\sum_b\eta_b=1$.

We assume that redshift is independent of masses and spins within each branch, while allowing $p_b(z|\lambda)$ to differ between branches.
This reflects the delay between the formation of 2G BHs in 1G+1G mergers and their subsequent mergers.
For each 1G BH we
assume its mass and spin to have a factorized distribution
\begin{equation}\label{eq: 1G distribution}
    \pOneG{m,\chi|\lambda} = p_m(m|\lambda)\,p_\chi(\chi|\lambda)\,.
\end{equation}
A 2G BH is the remnant of a 1G+1G merger, so that its mass and spin follow
\begin{equation}\label{eq: 2G distribution}
    p_{2\mathrm{G}}(m,\chi|\lambda) = \int \dd{\theta} \delta\big(m-M_f(\theta)\big) \delta\big(\chi-\chi_f(\theta)\big) p_{11}(\theta|\lambda) \, ,
\end{equation}
where $M_f(\theta)$ and
$\chi_f(\theta)$ are the remnant mass and spin, respectively, as functions of the progenitor parameters $\theta$.
The delta functions in Eq.~\eqref{eq: 2G distribution} encode the relation between the progenitor and the post-merger mass $M_f$ and spin $\chi_f$ and
integrating them out requires solving $M_f(\theta)=m$ and $\chi_f(\theta)=\chi$ for the progenitor parameters.
Since the remnant properties are nonlinear fits to numerical-relativity simulations that cannot be inverted analytically,
Eq.~\eqref{eq: 2G distribution} has no closed form and reduces at best to a five-dimensional integral that has to be computed numerically.
This integral is needed on every PE sample of every event at every likelihood call, and twice for the 2G+2G branch,
which makes evaluating the hierarchical-merger population extremely challenging.  In contrast, drawing samples from Eq.~\eqref{eq: 2G distribution} only requires a forward simulation, which is straightforward to implement and fast to run.

\subsection{Implementation details}

We now turn to the details of the mixture model ingredients:
\begin{enumerate}[leftmargin=*]
\item \textit{Remnant properties.}
To model the remnant mass and spin, we adopt the fits to NR simulations of quasicircular BH mergers of Refs.~\cite{Barausse:2012qz, Hofmann:2016yih}.
Implicitly, we neglect the dependence of the retention probability on the progenitor parameters
through the recoil of the remnant~\cite{Zevin:2022bfa,Gerosa:2019zmo}:
Eq.~\eqref{eq: 2G distribution} includes all remnants, and retention only enters through the branching ratios.
\item \textit{Pairing function.}
When sampling the two components of a binary, we adopt the same formalism as Refs.~\cite{Fishbach:2019bbm, Farah:2021qom, Farah:2023swu}
and pair them by means of a pairing function $\mathcal{P}_b(q)$, which depends on the branch $b$ and the mass ratio $q=m_2/m_1\leq1$,
\begin{equation}\label{eq: pairing function}
    \mathcal{P}_b(q) = q^{\beta_{q,b}}\,\Theta(q-q_{\rm min})\,,
\end{equation}
with $q_{\rm min}=0.1$ and one exponent $\beta_{q,b}$ per branch, so that 2G BHs can select their partners differently from 1G BHs.
Positive (negative) values of $\beta_{q,b}$ favor (disfavor) equal-mass pairings.
In practice, when sampling a binary, we draw the two components independently
from their respective distributions, take the heavier one as the primary, and accept the binary with probability
proportional to Eq.~\eqref{eq: pairing function}.
We note that this procedure allows for the primary to be either 1G or 2G, without forcing the 2G component to be the primary.
\item \textit{1G mass distribution.}
For the 1G mass distribution, we adopt a broken power law with a low-mass taper~
between $m_{\rm min}$ and $m_{\rm max}$,
\begin{equation}\label{eq: BPL 1G mass}
    \begin{gathered}
    p_m(m|\lambda) \propto S(m|m_{\rm min},\delta_m)\left(\frac{m}{m_{\rm break}}\right)^{-\alpha(m)},\\
    \alpha(m) =
    \begin{cases}
        \alpha_1\,, & m< m_{\rm break}\,,\\
        \alpha_2\,, & m\geq m_{\rm break}\,,
    \end{cases}
    \end{gathered}
\end{equation}
where $S$ is a Planck taper rising from 0 at $m_{\rm min}$ to 1 at $m_{\rm min}+\delta_m$~\cite{Talbot:2018cva}.
In contrast to the \defaultBBH model of Ref.~\cite{LIGOScientific:2026ctl}, Eq.~\eqref{eq: BPL 1G mass} has no Gaussian peaks
and describes each 1G component rather than the primary mass, as in Refs.~\cite{Fishbach:2019bbm, Farah:2023swu}.
We do not include a peak at $\sim\!35\,M_\odot$ because, with GWTC-5.0, this feature is consistent with a change in slope of both the primary and secondary mass distributions
\cite{LIGOScientific:2026ctl}.
The cutoff $m_{\rm max}$ applies to 1G BHs only, while 2G remnants can extend up to $2\,m_{\rm max}$.
\item \textit{Spin distribution.}
We model the spin distribution of the 1G BHs after Ref.~\cite{LIGOScientific:2025pvj}.
In particular, the spin magnitudes follow a Gaussian $p_\chi$ truncated to $[0,1]$, with mean $\mu_\chi$ and width $\sigma_\chi$, while the cosines of the tilts
follow independent Gaussians $p_{\cos\theta}$ truncated to $[-1,1]$, with mean $\mu_{\cos\theta}$ and width $\sigma_{\cos\theta}$.
The spin magnitudes of the 2G remnants are instead determined by the remnant fits
from (i),
while in binaries containing a 2G BH we assume the tilts of both components to be isotropic, as expected for dynamically formed binaries~\cite{Gerosa:2017kvu, Belczynski:2007xg, Rodriguez:2016vmx}.
\item \textit{Redshift distribution.}
Since 2G BHs form in mergers of 1G BHs, the redshift distribution must account for the delay between
the formation of each 2G BH and its subsequent merger.
We describe the 1G+1G mergers with the usual merger rate per unit comoving volume and source-frame time~\cite{Fishbach:2018edt},
\begin{equation}\label{eq: 11 rate}
    \mathcal{R}_{11}(z)\propto(1+z)^{\kappa_z}\,.
\end{equation}
Each 2G BH merges again after a delay $t_p$, measured from the 1G+1G merger that formed it at redshift $z$,
i.e., at lookback time $t_L(z)$, with~\cite{Gerosa:2017kvu}
\begin{equation}\label{eq: lookback time}
    t_L(z) = \frac{1}{H_0}\int_0^{z}\frac{\dd{z'}}{(1+z')\sqrt{\Omega_M(1+z')^3+\Omega_\Lambda}}\,,
\end{equation}
where $H_0=67.66\,\mathrm{km\,s^{-1}\,Mpc^{-1}}$ is the Hubble constant and $\Omega_M=0.31$ and $\Omega_\Lambda=1-\Omega_M$ are the matter and dark-energy density parameters of the Planck 2018 cosmology~\cite{Planck:2018vyg}.
We assume $t_p$ log-uniformly distributed in $[10^{-4}\,\mathrm{Gyr},\,t_L(z)]$~\cite{Gerosa:2017kvu},
so that the second merger happens at lookback time $\tau=t_L(z)-t_p\geq0$, i.e., before today.
In a 2G+2G binary, the delay is measured from the later of the two formation times, since the binary can only form once both 2G BHs exist.
The delays shift the merger rates $\mathcal{R}_{12}$ and $\mathcal{R}_{22}$ toward lower redshifts than $\mathcal{R}_{11}$,
and all three rates are converted into redshift distributions in the same way,
\begin{equation}\label{eq: branch redshift}
    p_b(z|\lambda) \propto \frac{\dd V_c}{\dd z}\,\frac{\mathcal{R}_b(z)}{1+z}\,,
\end{equation}
for $z\leq z_{\rm max}=3$, where $V_c$ is the comoving volume and the factor $1/(1+z)$ converts source-frame into observer-frame time.
We assume that 2G BHs merge again in the same environment, so that their number per unit comoving volume is conserved during the delay.
The rates $\mathcal{R}_{12}$ and $\mathcal{R}_{22}$ are then obtained by convolving the formation rate of their 2G components with the delay distribution $p(t_p|\tau)$,
\begin{align}
    \mathcal{R}_b(\tau) &\propto \int_{\tau+t_{\rm min}}^{t_L(z_{\rm max})} \dd{\tau'}\, r_b(\tau')\, p(t_p|\tau)\,,
    \notag\\
    r_{12}(\tau) &\propto \mathcal{R}_{11}(\tau)\,,
    \notag\\
    r_{22}(\tau) &= 2\,r_{12}(\tau)\left[1-\int_{t_{\rm min}}^{\tau}\dd{\tau'}\,r_{12}(\tau')\right]\,.
    \label{eq: delayed rate}
\end{align}
For $b=12$, the binary contains a single 2G BH, so that $r_{12}$ follows the 1G+1G merger rate.
For $b=22$, the delay starts at the later of two independent formation times, so that $r_{22}$ is the distribution of the minimum of two draws from $r_{12}$:
the term in square brackets is the probability that the other 2G BH formed earlier, and the factor 2 accounts for either of the two being the later one.

To evaluate Eq.~\eqref{eq: branch redshift}, we tabulate the three distributions on 2000 equal-width bins in $[0,z_{\rm max}]$.
Since Eq.~\eqref{eq: hierarchical mixture} factorizes in $z$ within each branch, the hybrid estimator of Eq.~\eqref{eq: pdet hybrid estimator} still applies, with each population sample reweighted by the tabulated $p_b(z|\lambda)$ of its own branch.

\end{enumerate}

\begin{table}[t]
\caption{Priors on the hyperparameters of the hierarchical-merger model of Sec.~\ref{sec: hierarchical mergers}. Here $\mathcal{U}(a,b)$ denotes a uniform distribution on $[a,b]$. Masses are in the source frame and in units of $M_\odot$. The priors are further restricted to $m_{\rm min}<m_{\rm break}<m_{\rm max}$.}
    \label{tab: hierarchical priors}
    \centering
    \begin{tabular}{lll}
    \toprule
    Parameter & Description & Prior \\
    \midrule
    \multicolumn{3}{l}{\textit{1G masses}} \\
    $\alpha_1$ & Slope below $m_{\rm break}$ & $\mathcal{U}(-4,12)$ \\
    $\alpha_2$ & Slope above $m_{\rm break}$ & $\mathcal{U}(-4,12)$ \\
    $m_{\rm break}$ & Break mass & $\mathcal{U}(10,60)$ \\
    $m_{\rm min}$ & Minimum mass & $\mathcal{U}(3,10)$ \\
    $m_{\rm max}$ & Maximum 1G mass & $\mathcal{U}(40,200)$ \\
    $\delta_m$ & Width of the low-mass taper & $\mathcal{U}(0,10)$ \\
    \midrule
    \multicolumn{3}{l}{\textit{1G spins}} \\
    $\mu_\chi$ & Mean of $\chi_i$ & $\mathcal{U}(0,1)$ \\
    $\sigma_\chi$ & Width of $\chi_i$ & $\mathcal{U}(0.005,1)$ \\
    $\mu_{\cos\theta}$ & Mean of $\cos\theta_i$ & $\mathcal{U}(-1,1)$ \\
    $\sigma_{\cos\theta}$ & Width of $\cos\theta_i$ & $\mathcal{U}(0.01,4)$ \\
    \midrule
    \multicolumn{3}{l}{\textit{Pairing and branching ratios}} \\
    $\beta_{q,b}$ & Pairing exponent of branch $b$ & $\mathcal{U}(-3,10)$ \\
    $\eta_{b}$ & Branching ratio of branch $b$ & Dirichlet \\
    \midrule
    \multicolumn{3}{l}{\textit{Redshift}} \\
    $\kappa_z$ & Power-law index of $\mathcal{R}_{11}(z)$ & $\mathcal{U}(-5,5)$ \\
    \bottomrule
    \end{tabular}
\end{table}

In total, this model has 16 hyperparameters, listed in Table~\ref{tab: hierarchical priors}, for which we take uniform priors,
except for the branching ratios, for which we take a Dirichlet prior.
When sampling at a given $\lambda$, we draw a reference set of $10^6$ binaries per branch and for each event keep $N_{\lambda, i}=1.2\times10^5$ of those falling within each $R_i$ (Sec.~\ref{sec: variance and restricted sampling}), according to the branching ratios $\eta_b$. The selection function is estimated again with Eq.~\eqref{eq: pdet hybrid estimator} on $N_\lambda=10^6$ binaries drawn from the full population, and we adopt the same sampler settings as in Sec.~\ref{sec: comparison}.

\subsection{Inference results}\label{sec: hierarchical mergers inference results}

\begin{figure*}
    \centering
    \includegraphics[width=\textwidth]{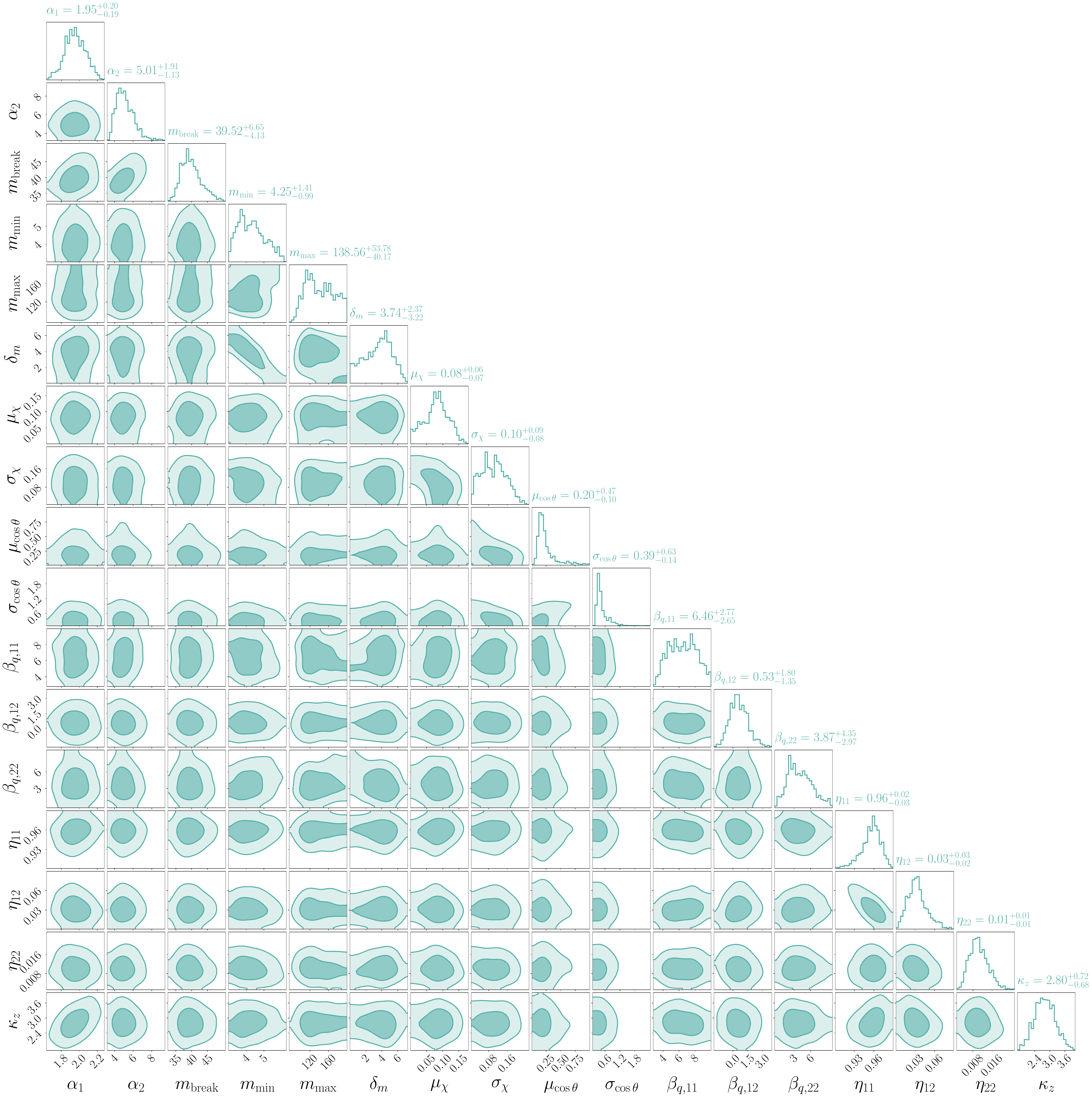}
    \caption{Posterior distribution of the population parameters of the hierarchical-merger model inferred from GWTC-5.0.
    Contours enclose the 50\% and 90\% credible regions; labels on the diagonal plots report medians and 90\% credible intervals.
    }
    \label{fig: gwtc-5 hierarchical inference corner}
\end{figure*}

\begin{figure*}[!t]
    \centering
    \includegraphics[width=\textwidth]{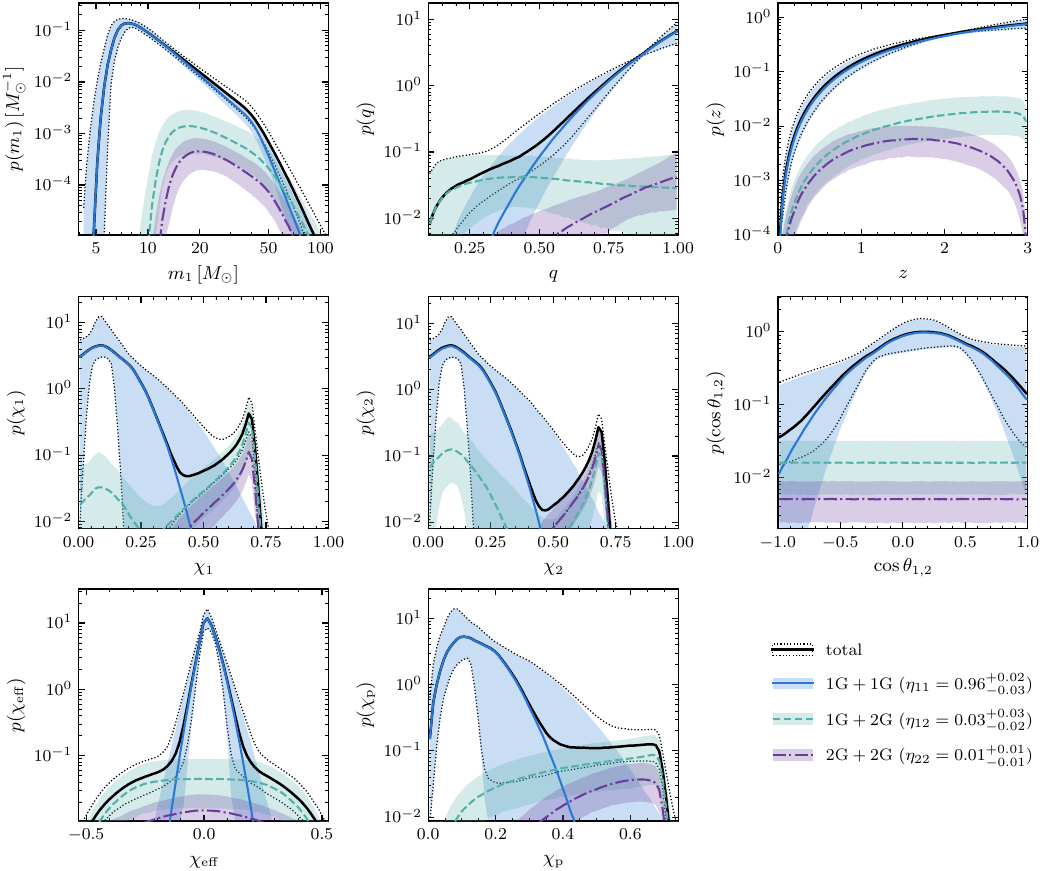}
    \caption{Inferred population distribution of the hierarchical-merger model of Sec.~\ref{sec: hierarchical mergers}
    with the three branches shown separately: 1G+1G in blue, 1G+2G in green, 2G+2G in purple, and their sum in black.
    The bands show the 90\% credible intervals computed from posterior samples, while the solid curves refer to the medians.
    In the top legend we report the branching ratios $\eta_b$ inferred for each branch, with the 90\% credible intervals.
    In addition to the parameters $\theta=(m_1,q,z,\chi_1,\chi_2,\cos\theta_1,\cos\theta_2)$ used in the inference, we show the distributions of the effective spins
    $\chi_{\rm eff}$ and $\chi_{\rm p}$ computed from the sampled parameters.
    }
    \label{fig: branch ppc}
\end{figure*}
The results of the inference are reported in Figs.~\ref{fig: gwtc-5 hierarchical inference corner} and~\ref{fig: branch ppc}, where
we show, respectively, the posterior of the hyperparameters and the inferred population distribution of the hierarchical-merger model with the three branches shown separately.

The inferred branching ratios show that the population is dominated by 1G+1G mergers, with $\eta_{11}=0.96^{+0.02}_{-0.03}$ (medians and 90\% credible intervals).
Hierarchical mergers constitute the remaining few percent, with $\eta_{12}=0.03^{+0.03}_{-0.02}$ and $\eta_{22}=0.01^{+0.01}_{-0.01}$.
The 90\% credible interval of $\eta_{12}$ excludes zero, whereas $\eta_{22}$ is consistent with zero.
A nonzero $\eta_{12}$ shows that the model uses the 1G+2G branch to describe a subset of the events, namely those with unequal masses, as discussed below.

This further provides an a posteriori justification for the broken power law of Eq.~\eqref{eq: BPL 1G mass}: being the dominant contribution, the 1G+1G
mass distribution must be similar to that of the full population inferred by the LVK Collaboration~\cite{LIGOScientific:2026ctl},
whose main features are reasonably described by a broken power law.
We further note that, compared to Refs.~\cite{LIGOScientific:2025pvj, LIGOScientific:2026ctl},
the peak at $\sim\!10\,M_\odot$ is partially absorbed by the low-mass taper of the broken power law,
thus resulting in a slight shift toward lower values and a shallower slope, as we do not fit specifically for that feature.

Regarding the mass-ratio distribution, we find that each branch pairs its components differently.
First-generation binaries strongly favor equal masses, $\beta_{q,11}=6.5^{+2.8}_{-2.7}$, so that the mass-ratio distribution of this branch drops by more
than an order of magnitude between $q=1$ and $q\approx0.5$ (Fig.~\ref{fig: branch ppc}).
For 1G+2G binaries, $\beta_{q,12}=0.5^{+1.8}_{-1.3}$ is instead consistent with random pairing ($\beta_{q,12}=0$), and the
distribution $p(q)$ appears approximately flat for $0.2\lesssim q\leq1$.
The 2G+2G branch again favors comparable masses, $\beta_{q,22}=3.9^{+4.3}_{-3.0}$, with a broader posterior because this branch contains only $\sim\!1\%$ of the binaries.
These pairing functions determine which branch a binary with a given mass ratio most likely belongs to.

\begin{figure}
    \centering
    \includegraphics[width=0.8\columnwidth]{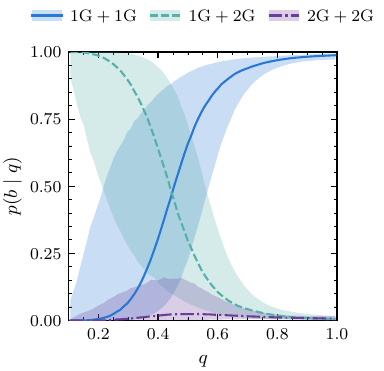}
    \caption{Fraction of binaries belonging to each branch as a function of the mass ratio $q$, $\eta_b\,p_b(q|\lambda)/\ppop(q|\lambda)$, for the hierarchical-merger model of Sec.~\ref{sec: hierarchical mergers}.
    Solid curves and shaded areas show medians and 90\% credible intervals.
    Equal-mass binaries ($q\sim1$) are dominated by 1G+1G mergers, while mergers involving 2G BHs are more common at low mass ratios.}
    \label{fig: branch fractions mass ratio}
\end{figure}

Figure~\ref{fig: branch fractions mass ratio} shows the fraction of binaries that belong to a given branch as a function of $q$, $\eta_b\,p_b(q|\lambda)/\ppop(q|\lambda)$, where $p_b(q|\lambda)$ and $\ppop(q|\lambda)$ are the mass-ratio marginals of branch $b$ and of the full population.
Near-equal-mass binaries are almost exclusively 1G+1G mergers: the median 1G+1G fraction exceeds 90\% for $q\gtrsim0.7$.
Below $q\approx0.45$, the 1G+2G branch becomes the most likely, because the steep 1G+1G pairing function suppresses unequal-mass first-generation binaries.
Events with posterior support at small mass ratios are therefore preferentially associated with hierarchical mergers.
The credible bands at $q\lesssim0.4$ are, however, broad, since only a few events constrain this region, and this association remains uncertain.

The drop at $z\sim3$ in the 2G redshift distributions of Fig.~\ref{fig: branch ppc} is an artifact of our implementation, which does not account for the
2G BHs formed at $z>z_{\rm max}$ and merging at $z<z_{\rm max}$.
By accounting for the fraction of those binaries that would be detectable by current detectors, we
estimate that this truncation might underestimate $\eta_{12}$ by $\lesssim\!5\%$ and $\eta_{22}$ by $\approx\!40\%$, which is still contained within their statistical uncertainty (90\% credible interval).
Given the illustrative purposes of this analysis, we defer a more refined investigation to future work.

\begin{figure*}[!t]
    \centering
    \includegraphics[width=\textwidth]{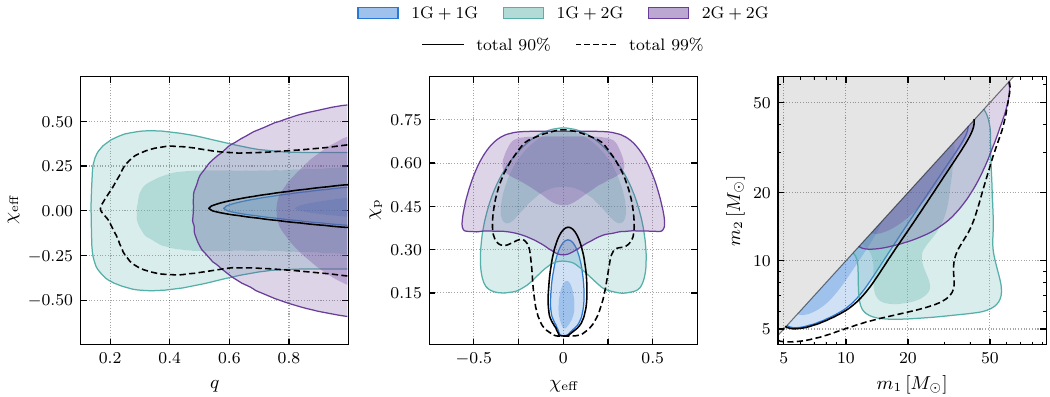}
    \caption{Joint distributions of pairs of binary parameters for the hierarchical-merger model of Sec.~\ref{sec: hierarchical mergers}.
    \textit{Left:} effective aligned spin $\chi_{\rm eff}$ versus mass ratio $q$. \textit{Center:} effective precession spin $\chi_{\rm p}$ versus $\chi_{\rm eff}$.
    \textit{Right:} source-frame primary mass $m_1$ versus source-frame secondary mass $m_2$.
    Shaded regions contain 90\% (darker) and 99\% (lighter) of the binaries in each branch: 1G+1G in blue, 1G+2G in green, and 2G+2G in purple.
    Solid and dashed black contours contain 90\% and 99\% of the full population.
    The effective spins are computed from the sampled parameters using Eqs.~\eqref{eq: chi eff} and~\eqref{eq: chi p}.}
    \label{fig: branch correlations}
\end{figure*}

The spin magnitudes carry the clearest signature of the hierarchical subpopulations.
The 1G BHs are slowly spinning, $\mu_\chi=0.08^{+0.06}_{-0.07}$ and $\sigma_\chi=0.10^{+0.09}_{-0.08}$, so that the 1G+1G contribution to $p(\chi_i)$ is negligible above $\chi_i\approx0.5$.
At larger values, $p(\chi_i)$ is dominated by the 2G BHs, whose spin distribution peaks at $\chi_i\approx0.7$~\cite{Fishbach:2017dwv, Gerosa:2017kvu} and drops sharply above it.

Figure~\ref{fig: branch ppc} also shows the distributions of two effective spin parameters that have been shown to be strong indicators of hierarchical mergers~\cite{Baibhav:2021qzw,Plunkett:2026pxt}.
The effective aligned spin~\cite{Damour:2001tu}
\begin{equation}\label{eq: chi eff}
    \chi_{\rm eff} = \frac{\chi_1\cos\theta_1 + q\,\chi_2\cos\theta_2}{1+q}
\end{equation}
is the mass-weighted projection of the spins onto the orbital angular momentum, which is approximately conserved during the inspiral~\cite{Racine:2008qv, Gerosa:2015tea}.
The effective precession spin~\cite{Schmidt:2014iyl} (but see Refs.~\cite{Thomas:2020uqj,Gerosa:2020aiw})
\begin{equation}\label{eq: chi p}
    \chi_{\rm p} = \max\left\{\chi_1\sin\theta_1,\ q\,\frac{4q+3}{4+3q}\,\chi_2\sin\theta_2\right\}
\end{equation}
measures instead the in-plane spin components, which drive the precession of the orbital plane.
The 1G+1G binaries form a narrow peak at $\chi_{\rm eff}\approx0$, slightly shifted toward positive values because the 1G tilts mildly favor alignment, $\mu_{\cos\theta}=0.20^{+0.47}_{-0.10}$, and have $\chi_{\rm p}\lesssim0.4$.
The hierarchical branches instead produce broad, flat tails in $\chi_{\rm eff}$, symmetric about zero, which dominate the population for $\chi_{\rm eff}\lesssim-0.15$ and $\chi_{\rm eff}\gtrsim0.2$.
This shape follows from the isotropic tilts of these binaries: the projection of the $\chi\approx0.7$ spin of the 2G component onto the orbital angular
 momentum is equally likely to take any value between $-0.7$ and $0.7$, and the mass weighting in Eq.~\eqref{eq: chi eff}
 narrows this range to $|\chi_{\rm eff}|\lesssim0.5$ for the typical $q\approx0.5$ of 1G+2G binaries.
Similarly, $p(\chi_{\rm p})$ is nearly flat for $0.4\lesssim\chi_{\rm p}\lesssim0.7$, where it is dominated by the hierarchical branches.
For isotropic tilts the distribution of $\sin\theta_i$ peaks at unity, so that $\chi_{\rm p}$ accumulates just below the 2G spin magnitude and drops sharply at $\chi_{\rm p}\approx0.7$, the same cutoff found in $p(\chi_i)$.
In this model, negative $\chi_{\rm eff}$ and large $\chi_{\rm p}$ are therefore signatures of hierarchical mergers.

We finally note that the cutoff at $\chi_i\approx0.7$ is an intrinsic property of the model rather than a feature resolved by the data.
Since the progenitors of 2G BHs are assumed to follow the same spin distribution as the 1G+1G binaries, the model assigns negligible probability to $\chi_i\gtrsim0.75$.
BHs produced by hierarchical mergers struggle to account for highly spinning components $\chi\sim 0.9$,
such as those of GW231123~\cite{LIGOScientific:2025rsn}.
Within this framework, such spins would require rapidly spinning 1G progenitors or mergers of higher generations, neither of which is included in our model.
This raises the question of whether this event has a hierarchical origin~\cite{Passenger:2025acb, Paiella:2025qld}.
Several alternative explanations have also been proposed~\cite{Croon:2025gol,Ray:2025rtt,Bartos:2025pkv}, alongside the possibility of uncontrolled waveform systematics~\cite{LIGOScientific:2025rsn}.
A comprehensive astrophysical interpretation of this event and of the entire catalog is, however, beyond the scope of this work.

We conclude by mentioning that one of the advantages of our method is that simulated models naturally provide multidimensional correlations between the parameters of each binary,
which are difficult to capture with parametric or nonparametric models.
We show some of these correlations in the three panels of Fig.~\ref{fig: branch correlations}, where we plot the joint distributions of
$(\chi_{\rm eff},q)$, $(\chi_{\rm p},\chi_{\rm eff})$, and source-frame component masses $(m_1,m_2)$ for the three branches separately.
We observe a slight broadening of the $\chi_{\rm eff}$ distribution at low $q$ for the 1G+2G branch.
This is in agreement with recent analyses~\cite{LIGOScientific:2025pvj,LIGOScientific:2026ctl}, which weakened the evidence
for a correlation in the mean $(\chi_{\rm eff},q)$~\cite{KAGRA:2021duu,Callister:2021fpo} in favor of a correlation in the width of $\chi_{\rm eff}$ with $q$.
Regarding the $(\chi_{\rm eff},\chi_{\rm p})$ joint distribution, the elliptic structure observed in the 2G branches is a key signature of hierarchical mergers,
derived analytically in Ref.~\cite{Baibhav:2021qzw} and then identified in GWTC-4.0 data by Ref.~\cite{Plunkett:2026pxt}.
In the $(m_1,m_2)$ panel of Fig.~\ref{fig: branch correlations}, we observe that the 1G+2G branch instead populates the in-between region. Due to this, we do not observe gaps in either primary or secondary mass distributions
which can be attributed to PISN~\cite{Belczynski:2016jno,Woosley:2016hmi,Farmer:2019jed},
but whose evidence is still debated~\cite{Ray:2025xti, Antonini:2025ilj, Tong:2025wpz,Mould:2026sww}.

Identifying the presence of hierarchical mergers in the population has attracted increasing attention as the number of detections grows.
We applied our method to this problem with a model that considers realistic population features,
like the relations linking the properties of each remnant to those of its progenitors.
Our results are overall consistent with previous findings~\cite{Gerosa:2017kvu,Doctor:2019ruh, Kimball:2020qyd,Mould:2022ccw,Wang:2022gnx,Ray:2026uur,Plunkett:2026pxt,Farah:2026jlc,Vijaykumar:2026zjy}.
We stress, however, that our model is not intended as a fully realistic description of the astrophysical population, but rather an illustrative application of our method.

\section{Population-synthesis models}\label{sec: pop synth}

Finally, we turn to population inference from astrophysical simulations, which is the driving science case for the methods developed in this paper.

\subsection{Grid of simulated catalogs}\label{sec: pop synth model grid}

Population-synthesis codes implement a set of astrophysical recipes to predict the properties of merging BHs from their stellar progenitors. In statistical terms, they are a forward model, i.e., they provide a finite set of sampled binaries; a closed-form density expression is not available, as they account
for the complex physics of stellar evolution and binary interactions~\cite{Spera:2015vkd,Breivik:2019lmt,Rodriguez:2021qhl,Fragos:2022jik,Kritos:2022ggc,COMPASTeam:2021tbl}.
These simulations are typically expensive to run, especially when they rely on detailed binary-evolution or stellar-cluster calculations,
and are performed on a coarse grid of a few physical parameters.
Our method only requires samples from $\ppop(\theta|\lambda)$ and could, in principle,
be used to run an end-to-end hierarchical inference, provided that the codes are fast enough to
draw a new set of simulated binaries
at every value of $\lambda$ proposed by the sampler.
However, the prohibitive cost of each simulation with current population-synthesis codes constitutes the main limitation of our approach.
Still, our method enables a direct comparison of the available simulated catalogs with the observed GW events,
via the computation of the population likelihoods at different grid points, without requiring any approximate density estimate of the population.

As a test case, we use the BH simulations of Ref.~\cite{Zevin:2020gbd}, which are publicly available~\cite{zevin_2021_4947741}, and tailor our analysis to their specific features.
We emphasize that this is an illustrative example: our approach is general and can be systematically applied to other simulation catalogs and used to compare them.
We plan to explore these broader applications in the near future.

\subsection{Implementation details}\label{sec: pop synth model}

The model of Ref.~\cite{Zevin:2020gbd} accounts for five different formation channels: the three isolated-binary evolution channels of common-envelope evolution (CE) and stable mass transfer (SMT) simulated using \textsc{\small POSYDON}~\cite{Fragos:2022jik, Bavera:2020inc}, and chemically homogeneous evolution (CHE) using MESA~\cite{duBuisson:2020asn}; and two cluster channels, globular clusters (GCs) simulated with \textsc{\small CMC}~\cite{Rodriguez:2021qhl, Rodriguez:2019huv} and a semianalytic model of nuclear star clusters (NSCs)~\cite{Antonini:2018auk}, both including hierarchical mergers.

Each of these channels is simulated by varying two physical parameters on a grid: the natal BH spin $\chi_b$ and common-envelope efficiency $\alpha_{\rm CE}$.
The natal spin $\chi_b\in\{0,0.1,0.2,0.5\}$ is the spin magnitude of BHs that are
born in isolation or in systems where binary interactions do not cause significant spin up,
and acts as a proxy for the efficiency of angular-momentum transport in massive stars~\cite{Zevin:2020gbd}.
It applies to all channels, including the first-generation BHs of the cluster channels.
The common-envelope efficiency $\alpha_{\rm CE}\in\{0.2,0.5,1,2,5\}$, on the contrary, only affects the CE channel and parametrizes how efficiently the orbital
energy is used to eject the envelope.
The CE channel is simulated at all 20 combinations of $(\chi_b,\alpha_{\rm CE})$, whereas the other four channels depend only on $\chi_b$
and are simulated at its four values, for a total of $20+4\times4=36$ simulations, each containing $\mathcal{O}(10^5$--$10^6)$ binaries.

Each simulated binary further carries a weight, proportional to the number of astrophysical mergers it represents~\cite{Zevin:2020gbd}.
The weights are needed because the initial conditions of the simulations, such as progenitor masses, orbital periods, metallicities, and formation times,
are sampled either on grids or from distributions chosen for computational convenience, rather than from their astrophysical distributions.
For the isolated channels, these weights account for the distribution of the progenitors' initial conditions
and for the metallicity-dependent star-formation history of Ref.~\cite{Madau:2016jbv}, which sets the formation redshift of each progenitor:
each weight is then the contribution of the binary to the observer-frame merger rate, in $\mathrm{yr^{-1}}$.
For the cluster channels, the weights account for the metallicities and formation times of the host clusters.

Following Ref.~\cite{Zevin:2020gbd}, we model the population as a mixture of the five channels,
\begin{equation}\label{eq: channel mixture}
    \ppop(\theta|\lambda) = \sum_{c\in\mathcal{C}} \eta_c\, p_c(\theta|\chi_b,\alpha_{\rm CE})\,,
\end{equation}
where $\mathcal{C}=\{{\rm CE},{\rm SMT},{\rm CHE},{\rm GC},{\rm NSC}\}$ and $\eta_c\geq0$ are the channel fractions, with $\sum_c\eta_c=1$.
The hyperparameters are therefore
\begin{equation}\label{eq: channel hyperparameters}
    \lambda = \{\chi_b,\alpha_{\rm CE},\eta_{\rm CE},\eta_{\rm SMT},\eta_{\rm CHE},\eta_{\rm GC},\eta_{\rm NSC}\}\,,
\end{equation}
where $(\chi_b,\alpha_{\rm CE})$ take the 20 discrete values of the simulations while the fractions are free parameters.
Drawing a binary from Eq.~\eqref{eq: channel mixture} amounts to choosing
the set corresponding to the specific $(\chi_b,\alpha_{\rm CE})$ grid point, then selecting a
channel $c$ with probability $\eta_c$
and then one of its binaries with probability proportional to $w_{ck}$.

Since Eq.~\eqref{eq: channel mixture} is linear in the $\eta_c$, so are the per-event integrals and the selection function.
Inserting it in Eq.~\eqref{eq: restricted population likelihood estimator} gives
\begin{align}
    \hat{\mathcal{L}}_R(\lambda) &= \hat{\xi}_{\rm GMM}(\lambda)^{-\Nobs}\prod_{i=1}^{\Nobs}\sum_{c\in\mathcal{C}}\eta_c\,\hat{\mathcal{L}}_{R_i,c}\,,
    \notag\\
    \hat{\xi}_{\rm GMM}(\lambda) &= \sum_{c\in\mathcal{C}}\eta_c\,\hat{\xi}_{{\rm GMM},c}\,,
    \label{eq: channel likelihood}
\end{align}
where
\begin{equation}\label{eq: channel event term}
    \hat{\mathcal{L}}_{R_i,c} = \frac{1}{\sum_{k=1}^{N_c}w_{ck}}\sum_{\theta_{ck}\in R_i} w_{ck}\,\frac{p(\theta_{ck}|d_i)}{\ppe(\theta_{ck})}
\end{equation}
is the restricted estimator [cf. Eq.~\eqref{eq: restricted population likelihood estimator}] of the likelihood of event $i$ for the population of channel $c$, and
for which
\begin{equation}
    P(R_i|\lambda) \simeq \sum_{c\in\mathcal{C}}\eta_c\left[ \frac{\sum_{\theta_{ck}\in R_i}w_{ck}}{\sum_{k}w_{ck}}\right]\,.
\end{equation}
Similarly, $\hat{\xi}_{{\rm GMM},c}$ is the selection-function estimator of Eq.~\eqref{eq: pdet gmm estimator} computed from samples of channel $c$.
Because the simulated binaries are fixed, restricting to $R_i$ does not reduce the Monte Carlo variance here: Eq.~\eqref{eq: channel event term} is identical to the unrestricted weighted average with $p(\theta|d_i)$ set to zero outside $R_i$.
We nonetheless retain this restriction, as it limits the GMM evaluations to the binaries within $R_i$ and prevents the fits from being extrapolated outside the posterior support.
Since $\hat{\mathcal{L}}_{R_i,c}$ and $\hat{\xi}_{{\rm GMM},c}$ do not depend on the fractions, we compute them once for all the simulations,
after which Eq.~\eqref{eq: channel likelihood} is an analytic function of the $\eta_c$ that involves only $5\,(\Nobs+1)$ precomputed numbers per grid point.

For GW231123~\cite{LIGOScientific:2025rsn}, none of the binaries in any of the 36 simulations fall within the event's region $R_i$.
Consequently, $\hat{\mathcal{L}}_{R_i,c}=0$ for every channel and grid point, and the likelihood in Eq.~\eqref{eq: channel likelihood} vanishes for every $\lambda$.
This largely reflects the history of the field: before the detection of GW231123, astrophysical simulations did not typically target such massive systems.
For this exploratory analysis, we exclude this event from the product in Eq.~\eqref{eq: channel likelihood} and perform the inference using the remaining 258 events of GWTC-5.0.
We discuss the implications of this choice in Sec.~\ref{sec: GW231123 exclusion}, where we argue that excluding this event does not significantly affect our results.

For each fixed combination of $(\chi_b,\alpha_{\rm CE})$, we sample the posterior for the four free fractions $\eta_c$ using \textsc{\small Dynesty}~\cite{Speagle:2019ivv}
through \textsc{\small Bilby}~\cite{Ashton:2018jfp}.
We set \texttt{nlive=500} and adopt a flat Dirichlet prior.
Each run returns the evidence
\begin{equation}\label{eq: grid evidence}
    \mathcal{Z}(\chi_b,\alpha_{\rm CE}) = \int \dd{\eta}\,\hat{\mathcal{L}}_R(\lambda)\,\pi(\eta)\,,
\end{equation}
with $\eta=\{\eta_c\}_{c\in\mathcal{C}}$ and $\pi(\eta)$ the Dirichlet prior, which can be further used to
perform model selection between the different simulations.
For this inference, we do not impose a cut on $\mathrm{Var}[\log\hat{\mathcal{L}}_R]$. We instead evaluate it, {including the cross terms of Eq.~\eqref{eq: log-likelihood variance full}}, at each posterior sample
from the sample variances of $\hat{\mathcal{L}}_{R_i,c}$ and $\hat{\xi}_{{\rm GMM},c}$, and report its posterior median at each grid point.

Note how we are comparing GW data and astrophysical models by directly using their simulation results, without population-synthesis interpolants as in Refs.~\cite{Zevin:2020gbd, Taylor:2018iat, Wong:2019uni, Wong:2020ise, Mould:2022ccw, Colloms:2025hib}.

\subsection{Inference results}
\label{sec: pop synth inference results}

\begin{table*}
    \centering
    \setlength{\tabcolsep}{4pt}
    \begin{tabular}{ccccc!{\hspace{0.5em}}ccccc}
        \toprule
         &  &  &  &  & \multicolumn{5}{c}{median $\eta_c$} \\
        \cmidrule(lr){6-10}
        $\chi_b$ & $\alpha_{\rm CE}$ & $\log\mathcal{Z}$ & $\Delta\log\mathcal{Z}$ & ${\rm Var}[\log\mathcal{L}]$ & ${\rm CE}$ & ${\rm SMT}$ & ${\rm CHE}$ & ${\rm GC}$ & ${\rm NSC}$ \\
        \midrule
        0 & 0.2 & $-778.1 \pm 0.2$ & \cellcolor[RGB]{255,243,203}$-58.6$ & \cellcolor[RGB]{255,254,221}$9.0$ & \cellcolor[RGB]{238,245,252}0.07 & \cellcolor[RGB]{116,179,216}0.68 & \cellcolor[RGB]{247,251,255}0.00 & \cellcolor[RGB]{231,241,250}0.12 & \cellcolor[RGB]{231,240,250}0.12 \\
        0 & 0.5 & $-765.7 \pm 0.2$ & \cellcolor[RGB]{255,247,210}$-46.2$ & \cellcolor[RGB]{245,251,209}$6.9$ & \cellcolor[RGB]{121,181,217}0.67 & \cellcolor[RGB]{214,229,244}0.24 & \cellcolor[RGB]{247,251,255}0.00 & \cellcolor[RGB]{242,247,253}0.04 & \cellcolor[RGB]{241,247,253}0.05 \\
        0 & 1 & $-758.1 \pm 0.2$ & \cellcolor[RGB]{255,250,215}$-38.7$ & \cellcolor[RGB]{237,248,199}$5.9$ & \cellcolor[RGB]{100,169,211}0.75 & \cellcolor[RGB]{224,236,248}0.17 & \cellcolor[RGB]{247,251,255}0.00 & \cellcolor[RGB]{242,248,253}0.04 & \cellcolor[RGB]{242,247,253}0.04 \\
        0 & 2 & $-729.0 \pm 0.2$ & \cellcolor[RGB]{227,243,191}$-9.6$ & \cellcolor[RGB]{246,251,211}$7.2$ & \cellcolor[RGB]{62,142,196}0.91 & \cellcolor[RGB]{241,247,253}0.05 & \cellcolor[RGB]{247,251,255}0.00 & \cellcolor[RGB]{245,250,254}0.01 & \cellcolor[RGB]{245,249,254}0.02 \\
        0 & 5 & $-719.5 \pm 0.2$ & \cellcolor[RGB]{138,198,165}$0.0$ & \cellcolor[RGB]{238,248,200}$6.0$ & \cellcolor[RGB]{60,140,195}0.93 & \cellcolor[RGB]{242,247,253}0.04 & \cellcolor[RGB]{247,251,255}0.00 & \cellcolor[RGB]{245,250,254}0.01 & \cellcolor[RGB]{245,249,254}0.02 \\
        \midrule
        0.1 & 0.2 & $-835.0 \pm 0.2$ & \cellcolor[RGB]{254,225,185}$-115.5$ & \cellcolor[RGB]{255,246,207}$11.4$ & \cellcolor[RGB]{242,248,253}0.04 & \cellcolor[RGB]{157,202,225}0.54 & \cellcolor[RGB]{247,251,255}0.00 & \cellcolor[RGB]{213,229,244}0.25 & \cellcolor[RGB]{223,236,247}0.17 \\
        0.1 & 0.5 & $-832.6 \pm 0.1$ & \cellcolor[RGB]{254,226,185}$-113.1$ & \cellcolor[RGB]{255,247,210}$10.9$ & \cellcolor[RGB]{202,221,240}0.33 & \cellcolor[RGB]{196,218,238}0.37 & \cellcolor[RGB]{247,251,255}0.00 & \cellcolor[RGB]{223,235,247}0.17 & \cellcolor[RGB]{231,240,250}0.12 \\
        0.1 & 1 & $-841.4 \pm 0.1$ & \cellcolor[RGB]{254,223,183}$-121.9$ & \cellcolor[RGB]{255,249,213}$10.3$ & \cellcolor[RGB]{185,214,234}0.42 & \cellcolor[RGB]{206,224,242}0.30 & \cellcolor[RGB]{247,251,255}0.00 & \cellcolor[RGB]{225,237,248}0.16 & \cellcolor[RGB]{231,241,250}0.11 \\
        0.1 & 2 & $-811.0 \pm 0.2$ & \cellcolor[RGB]{254,232,190}$-91.5$ & \cellcolor[RGB]{255,252,217}$9.6$ & \cellcolor[RGB]{77,153,202}0.84 & \cellcolor[RGB]{239,246,252}0.06 & \cellcolor[RGB]{247,251,255}0.00 & \cellcolor[RGB]{240,246,253}0.05 & \cellcolor[RGB]{242,247,253}0.04 \\
        0.1 & 5 & $-816.9 \pm 0.2$ & \cellcolor[RGB]{254,230,189}$-97.4$ & \cellcolor[RGB]{255,253,219}$9.4$ & \cellcolor[RGB]{66,146,198}0.89 & \cellcolor[RGB]{242,248,253}0.04 & \cellcolor[RGB]{247,251,255}0.00 & \cellcolor[RGB]{242,248,253}0.04 & \cellcolor[RGB]{243,248,254}0.03 \\
        \midrule
        0.2 & 0.2 & $-963.7 \pm 0.1$ & \cellcolor[RGB]{252,199,169}$-244.2$ & \cellcolor[RGB]{254,223,183}$17.9$ & \cellcolor[RGB]{242,247,253}0.04 & \cellcolor[RGB]{206,224,242}0.30 & \cellcolor[RGB]{247,251,255}0.00 & \cellcolor[RGB]{190,216,236}0.39 & \cellcolor[RGB]{211,228,243}0.26 \\
        0.2 & 0.5 & $-968.4 \pm 0.2$ & \cellcolor[RGB]{252,197,168}$-249.0$ & \cellcolor[RGB]{254,231,190}$15.6$ & \cellcolor[RGB]{178,210,232}0.45 & \cellcolor[RGB]{227,238,248}0.15 & \cellcolor[RGB]{247,251,255}0.00 & \cellcolor[RGB]{214,229,244}0.24 & \cellcolor[RGB]{225,237,248}0.16 \\
        0.2 & 1 & $-979.9 \pm 0.1$ & \cellcolor[RGB]{251,196,167}$-260.4$ & \cellcolor[RGB]{254,225,185}$17.5$ & \cellcolor[RGB]{211,228,243}0.26 & \cellcolor[RGB]{217,231,245}0.22 & \cellcolor[RGB]{247,251,255}0.00 & \cellcolor[RGB]{205,224,241}0.31 & \cellcolor[RGB]{219,233,246}0.20 \\
        0.2 & 2 & $-955.2 \pm 0.2$ & \cellcolor[RGB]{252,200,169}$-235.7$ & \cellcolor[RGB]{254,233,191}$15.2$ & \cellcolor[RGB]{95,166,209}0.77 & \cellcolor[RGB]{242,248,253}0.03 & \cellcolor[RGB]{247,251,255}0.00 & \cellcolor[RGB]{231,241,250}0.11 & \cellcolor[RGB]{236,244,251}0.08 \\
        0.2 & 5 & $-933.4 \pm 0.2$ & \cellcolor[RGB]{252,204,171}$-214.0$ & \cellcolor[RGB]{254,235,193}$14.8$ & \cellcolor[RGB]{75,152,202}0.85 & \cellcolor[RGB]{245,250,254}0.01 & \cellcolor[RGB]{247,251,255}0.00 & \cellcolor[RGB]{237,244,252}0.08 & \cellcolor[RGB]{240,246,253}0.06 \\
        \midrule
        0.5 & 0.2 & $-1167.0 \pm 0.2$ & \cellcolor[RGB]{246,175,158}$-447.5$ & \cellcolor[RGB]{242,166,154}$53.7$ & \cellcolor[RGB]{246,250,255}0.01 & \cellcolor[RGB]{242,248,253}0.04 & \cellcolor[RGB]{247,251,255}0.00 & \cellcolor[RGB]{100,169,211}0.74 & \cellcolor[RGB]{220,233,246}0.20 \\
        0.5 & 0.5 & $-1146.9 \pm 0.2$ & \cellcolor[RGB]{247,176,158}$-427.5$ & \cellcolor[RGB]{242,166,154}$53.6$ & \cellcolor[RGB]{236,244,251}0.08 & \cellcolor[RGB]{243,248,254}0.03 & \cellcolor[RGB]{247,251,255}0.00 & \cellcolor[RGB]{114,178,216}0.69 & \cellcolor[RGB]{221,234,247}0.19 \\
        0.5 & 1 & $-1161.4 \pm 0.2$ & \cellcolor[RGB]{246,175,158}$-442.0$ & \cellcolor[RGB]{242,166,154}$53.3$ & \cellcolor[RGB]{240,246,253}0.05 & \cellcolor[RGB]{242,248,253}0.03 & \cellcolor[RGB]{247,251,255}0.00 & \cellcolor[RGB]{108,174,214}0.71 & \cellcolor[RGB]{220,234,246}0.19 \\
        0.5 & 2 & $-1150.3 \pm 0.2$ & \cellcolor[RGB]{247,176,158}$-430.8$ & \cellcolor[RGB]{242,166,154}$52.1$ & \cellcolor[RGB]{226,237,248}0.15 & \cellcolor[RGB]{244,249,254}0.03 & \cellcolor[RGB]{247,251,255}0.00 & \cellcolor[RGB]{130,187,219}0.63 & \cellcolor[RGB]{223,235,247}0.17 \\
        0.5 & 5 & $-1150.1 \pm 0.2$ & \cellcolor[RGB]{247,176,158}$-430.6$ & \cellcolor[RGB]{242,166,154}$50.0$ & \cellcolor[RGB]{207,225,242}0.30 & \cellcolor[RGB]{245,249,254}0.02 & \cellcolor[RGB]{247,251,255}0.00 & \cellcolor[RGB]{160,203,226}0.53 & \cellcolor[RGB]{227,238,249}0.15 \\
        \bottomrule
        \end{tabular}
    \caption{Evidence of the five-channel mixture of Eq.~\eqref{eq: channel mixture} at each grid point $(\chi_b, \alpha_{\rm CE})$ of the models
    of Ref.~\cite{Zevin:2020gbd}. As a model selection criterion, we report the log-evidence $\log\mathcal{Z}$, with the statistical uncertainty estimated by \textsc{\small Dynesty}, and
    its difference $\Delta\log\mathcal{Z}$ with respect to the grid point with the largest evidence, $(\chi_b, \alpha_{\rm CE})=(0,5)$, computed from the unrounded values.
    The fifth column reports the posterior median of the Monte Carlo variance of the log-likelihood, $\mathrm{Var}[\log\hat{\mathcal{L}}_R]$, {including the covariances between the terms computed from the same table}.
    The last five columns report the posterior medians of the intrinsic fractions $\eta_c$. Cell colors scale with the value in each column.}
    \label{tab:zevin_grid_mixed}
    \end{table*}

\begin{figure*}
    \centering
    \includegraphics[width=\textwidth]{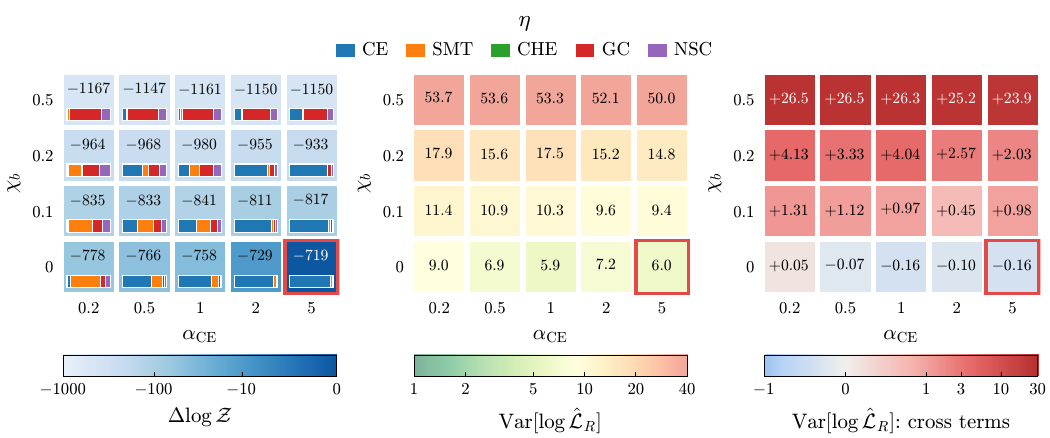}
    \caption{Alternative representation of the results in Table~\ref{tab:zevin_grid_mixed}.
    \textit{Left.} The boxes report the log-evidence $\log\mathcal{Z}$ at each grid point $(\chi_b, \alpha_{\rm CE})$, rounded to the nearest integer, while
    the color gradient encodes the log-evidence difference $\Delta\log\mathcal{Z}$ with respect to the grid point with the largest evidence, $(\chi_b, \alpha_{\rm CE})=(0,5)$.
    Within each box, the colored bars represent the posterior medians of the intrinsic fractions $\eta_c$ of the five channels, rescaled to sum to unity.
    \textit{Middle.} The boxes report the posterior median of the Monte Carlo variance of the log-likelihood, $\mathrm{Var}[\log\hat{\mathcal{L}}_R]$, at each grid point,
    also encoded by the color gradient.
    \textit{Right.} The boxes report the contribution of the cross terms to the total variance at each grid point, with the color gradient encoding their values. These include both covariances between the likelihood terms of different events and covariances between each single-event likelihood term and the selection-effect estimator.
    In all panels, red outlines mark the preferred grid point $(\chi_b=0,\alpha_{\rm CE}=5)$.
}
    \label{fig: zevin plot}
\end{figure*}

Results from this analysis are presented in Table~\ref{tab:zevin_grid_mixed} and Fig.~\ref{fig: zevin plot},
which contain the same information in different formats, for a complementary view of the results.
Table~\ref{tab:zevin_grid_mixed} lists the evidence, the median variance, and the median intrinsic fractions at each grid point.
To perform model selection, we report the log-evidence differences $\Delta\log\mathcal{Z}=\log\mathcal{Z}-\max(\log\mathcal{Z})$
with respect to the grid point with the largest evidence, i.e., the log Bayes factor between each grid point and the preferred one.

The data favor nonspinning natal BHs and efficient common-envelope ejection.
The largest evidence is obtained at $(\chi_b,\alpha_{\rm CE})=(0,5)$, followed by $(0,2)$ with $\Delta\log\mathcal{Z}=-9.6$,
while all the other grid points have $\Delta\log\mathcal{Z}\leq-38.7$.
At every $\alpha_{\rm CE}$, the evidence decreases monotonically with the natal spin, and every grid point with $\chi_b\geq0.1$ is disfavored by $\Delta\log\mathcal{Z}\leq-91.5$.

In the field channels, the spins are nearly aligned with the orbital angular momentum, so that a natal spin $\chi_b$
increases the effective spin [cf. Eq.~\eqref{eq: chi eff}] of most binaries to $\chi_{\rm eff}\gtrsim\chi_b$,
whereas the effective spins of GWTC-5.0 events concentrate near zero~\cite{LIGOScientific:2025pvj, LIGOScientific:2026ctl}.
At larger $\chi_b$, the inference compensates with the cluster channels, whose isotropic spins produce $\chi_{\rm eff}$ distributed symmetrically around zero:
at $\chi_b=0.5$, GC and NSC binaries indeed make up 68--94\% of the mergers, despite a much lower evidence.
At $\chi_b=0$, larger values of $\alpha_{\rm CE}$ are favored, likely because a more efficient ejection of the envelope leaves wider post-CE orbits,
in which the tidal spin-up of the second-born BH is weaker~\cite{Bavera:2020inc}.
At the preferred grid point, the intrinsic population is dominated by the CE channel, with median fraction $\eta_{\rm CE}=0.93$,
followed by SMT (0.04), NSC (0.02), and GC (0.01), while the CHE fraction is negligible.

The posterior median of the Monte Carlo variance of the log-likelihood ranges from 5.9 to 53.7 across the grid, well above the threshold $\mathrm{Var}[\log\hat{\mathcal{L}}]<4$
adopted in Sec.~\ref{sec: comparison}, which would exclude every grid point.
This is the main limitation due to the finite sample size of the catalogs.
In Secs.~\ref{sec: comparison} and~\ref{sec: hierarchical mergers}, the variance is controlled by drawing $N_{\lambda,i}\simeq10^5$ new population samples within each $R_i$ at every likelihood call.
Here, the set of binaries is fixed by the simulations, and Eq.~\eqref{eq: channel event term} only uses the binaries
that fall within $R_i$,
typically a small fraction of the $\mathcal{O}(10^5$--$10^6)$ available for events with well-measured masses, whose effective number is further reduced by the spread of the weights.
The variance grows with $\chi_b$, and it is correlated with the shift of the fractions toward the cluster channels.
This is due, again, to the smaller number of binaries in those channels, $(1.0$--$2.3)\times10^5$ compared with $10^6$ for CE and SMT,
which reduces the effective sample size.
The fixed set of simulations also implies that the
Monte Carlo errors are correlated and the cross-terms of Eq.~\eqref{eq: log-likelihood variance full} become relevant.
At $\chi_b=0$, where the field channels reproduce the bulk of the catalog, the covariances change the variance only by $-0.16$ to $+0.05$.
Their contribution grows with $\chi_b$, up to a point ($\chi_b=0.5$) where they become comparable to the diagonal ones.
The cross-variance terms must therefore be included in the variance of any analysis that uses the simulated binaries directly,
and they can be reduced with either more samples or even dedicated simulations for the selection function.

The Monte Carlo variance affects the results in two ways.
First, it sets the accuracy of the evidence at each grid point: the larger it is, the less accurate, and potentially more biased, the evidence estimate becomes~\cite{Talbot:2023pex}.
Second, whether a comparison between two grid points can be trusted depends on how their evidence difference compares with this uncertainty.
The grid points corresponding to the largest variance, at $\chi_b=0.5$, therefore have the least accurate evidences. At the same time,
they are disfavored by far more than their uncertainty,
so that the conclusion about the natal spins still holds.
On the contrary, the two most favored grid points, $(0,5)$ and $(0,2)$, have among the lowest variances, 6.0 and 7.2,
but their evidence difference, $\Delta\log\mathcal{Z}=-9.6$, is only a few times the corresponding Monte Carlo standard deviations of the log-likelihood, $\simeq2.5$ and $\simeq2.7$.
Therefore, the preference for $(0,5)$ over $(0,2)$ is less robust.

\subsection{Comparison with previous results}

These results are consistent with previous analyses of the same population-synthesis simulations~\cite{Zevin:2020gbd,Colloms:2025hib,Cheng:2023ddt}.
Using kernel density estimates and GWTC-2.0~\cite{LIGOScientific:2020kqk}, Ref.~\cite{Zevin:2020gbd} found $\chi_b\lesssim0.1$ favored over $\chi_b\gtrsim0.2$
by a Bayes factor of $\simeq13$ and reported a marginal preference for $\alpha_{\rm CE}=5$ over $\alpha_{\rm CE}=1$, with a Bayes factor of $\simeq5$.
Reference~\cite{Cheng:2023ddt} confirms those findings considering GWTC-3.0~\cite{KAGRA:2021duu} and, in particular,
it highlights the uncertainty in the inference of the branching ratios and the preference for $\chi_b=0$ over $\chi_b=0.1$ due to the limited number of simulations.
Reference~\cite{Colloms:2025hib} instead interpolated the simulations across $(\chi_b,\alpha_{\rm CE})$ by means of normalizing flows and, considering GWTC-3.0~\cite{KAGRA:2021duu},
inferred $\chi_b=0.04^{+0.04}_{-0.01}$, $\alpha_{\rm CE}>3.7$, and $\eta_{\rm CE}=0.91^{+0.05}_{-0.10}$ (medians and 90\% credible intervals),
with SMT, CHE, GC, and NSC fractions close to our inferred medians at $(0,5)$.
With the 258 binary BHs of GWTC-5.0, the preference for nonspinning natal BHs over $\chi_b=0.1$ becomes decisive.

We further highlight that both analyses excluded GW190521~\cite{LIGOScientific:2020iuh} as an outlier of these models,
whereas in our case only GW231123 is excluded.
GW190521, like GW231123, indeed lies in a region populated by only a few simulated binaries, where density estimates
built from the simulations struggle to accurately describe the population.
A strength of our method is that, on the contrary, it evaluates the fit of the event posterior directly on the simulated binaries,
so that even a few samples within its support give a nonzero, albeit noisy, contribution to the likelihood.

\subsection{On the exclusion of GW231123}\label{sec: GW231123 exclusion}
The exclusion of an event from the catalog can be treated as an additional selection effect,
which implies that the likelihood of Eq.~\eqref{eq: standard hierarchical likelihood}
remains consistent only if the same criterion is accounted for by the selection function~\cite{
Essick:2021vlx,Roy:2025ktr}.
An exclusion independent of the event parameters represents a random thinning of the Poisson process underlying
the events' realization, which in turn
rescales the merger rate as $\mathcal{R}\to(1-f)\,\mathcal{R}$, with $f$ the fraction of events excluded.
The same correction must be applied to the selection function, $\xi(\lambda)\to(1-f)\,\xi(\lambda)$.
In the case of O4, $f\simeq1/190\approx0.5\%$, which is negligible compared with the Monte Carlo uncertainty of the selection function.

In our specific case, however, the exclusion of GW231123 actually depends on where the event lies in parameter space.
Denoting by $R_\ast$ the region $R_i$ of this event, the exclusion multiplies the detection probability by $\Theta(\theta\notin R_\ast)$,
and thus removes from the selection function all the detectable binaries within $R_\ast$:
\begin{align}
    \xi(\lambda)&\to\xi(\lambda)\left[1-f_\ast(\lambda)\right]\,,
    \notag\\
    f_\ast(\lambda)&=\frac{1}{\xi(\lambda)}\int_{R_\ast}\dd{\theta}\,p(\det|\theta)\,\ppop(\theta|\lambda)\,,
    \label{eq: excluded detectable fraction}
\end{align}
where $f_\ast$ is the fraction of detectable binaries that lie within $R_\ast$.
Through the factor $\xi(\lambda)^{-\Nobs}$ of Eq.~\eqref{eq: standard hierarchical likelihood}, this cut increases the log-likelihood by
\begin{equation}\label{eq: cut log-likelihood shift}
    \Delta\log\mathcal{L}(\lambda) = -\Nobs\log\left[1-f_\ast(\lambda)\right] \simeq \Nobs\,f_\ast(\lambda)\,.
\end{equation}
In our restricted estimator, $f_\ast$ is the fraction of simulated binaries that fall within $R_\ast$.
Since there are none in the simulations, this cut leaves both the selection function and the terms of the other events unchanged.

This does not imply that the true $f_\ast$ is zero; the fraction may simply be too small to be resolved with the finite simulation samples.
If the true fraction of detectable binaries within $R_\ast$ is $f_\ast$, the probability that none of the $N$ effective binaries lies within this region is $(1-f_\ast)^N\simeq e^{-Nf_\ast}$.
This probability drops below 5\% when $Nf_\ast>3$.
For the simulations of Ref.~\cite{Zevin:2020gbd}, each table has an effective sample size of at least $1.3\times10^4$ after weighting by the detection probability, giving a 95\% upper limit of $f_\ast\lesssim2.3\times10^{-4}$ for each channel and hence for any mixture of channels.
Substituting this upper limit into Eq.~\eqref{eq: cut log-likelihood shift} gives an estimated upper bound of $\Delta\log\mathcal{L}\simeq0.06$ at every grid point.
This is much smaller than both the evidence differences and the Monte Carlo uncertainty reported in Table~\ref{tab:zevin_grid_mixed}.

This exclusion is, however, decided after seeing the data. 
The evidences in Table~\ref{tab:zevin_grid_mixed} therefore measure how well each grid point describes the binaries outside $R_\ast$,
but do not account for the fact that none of the 20 grid points can produce GW231123.

\section{Conclusions}\label{sec: conclusions}
Interpreting observations in terms of astrophysical models is a key goal of GW astronomy. In this work, we presented a step forward in this direction by developing a method for hierarchical population inference that directly exploits simulated binaries, without requiring density estimates or interpolants of the population model.

Our understanding of the mechanisms that form compact binaries is encoded in numerical population-synthesis simulations, which approximate the complex physics of stellar evolution and binary interactions and predict the properties of the resulting systems.
Directly using such simulated outputs in GW data analysis requires a statistical procedure that relies solely on samples rather than evaluations of the population-model probability density. We showed how this can be achieved by appropriately recasting the Monte Carlo estimators of the population likelihood, and applied the method to three cases: a standard parametric model (for validation purposes), a scenario involving hierarchical BH mergers, and a set of population-synthesis simulations from five different formation channels. 

Crucially, all the technical ingredients of this method---namely
surrogates for the single-event likelihoods and selection function, and our restricted-sampling strategy---do not depend on the specific population model but only on the GW data and detector network. These products can thus be constructed
and released once per GW observing run, and then used to test multiple astrophysical models.
This contrasts with other approaches that require interpolants of the population model itself,
which need to be retrained repeatedly~\cite{Zevin:2020gbd,Taylor:2018iat,Wong:2019uni,Wong:2020ise,Mould:2022ccw,Colloms:2025hib,Plunkett:2025mjr}.

When targeting a given set of astrophysical assumptions or mechanisms, running population-synthesis simulations involves
a trade-off between the density of points in the population-parameter space
(i.e., how many simulations should be run) and their statistical accuracy
(i.e., how many binaries should be generated in each simulation).
This is mainly due to the cost of running the simulations, which prevents one from generating either an arbitrarily
dense coverage of the parameter space or a large number of binaries at every grid point.
With the strategy presented here, the former determines the resolution of the resulting model selection, while the latter directly controls the Monte Carlo variance. Exploring this trade-off in realistic, large-scale population-synthesis studies will be an important step toward deploying the approach presented in this paper at scale.

\begin{acknowledgments}
We thank
Tristan Bruel, 
Cecilia Maria Fabbri, 
Stephen Green, 
Ana Lorenzo-Medina,
Michele Mancarella,
Stefano Rinaldi, 
Rodrigo Tenorio, 
and 
Alexandre Toubiana 
for discussions.
F.D.S. and D.G. are supported by
ERC Starting Grant No.~945155--GWmining,
Cariplo Foundation Grant No.~2021-0555,
Italian-French University (UIF/UFI) Grant No.~2025-C3-386,
MUR Grant ``Progetto Dipartimenti di Eccellenza 2023-2027'' (BiCoQ),
and the INFN TEONGRAV initiative.
D.G. is supported by MUR Young Researchers Grant No. SOE2024-0000125.
M.M. is supported by a Research Fellowship from the Royal Commission for the Exhibition of 1851.
Computational work was performed
at CINECA with allocations through INFN and the University of Milano-Bicocca.
This research has made use of data or software obtained from the Gravitational Wave Open Science Center (gwosc.org), a service of the LIGO Scientific Collaboration, the Virgo Collaboration, and KAGRA, and is based upon work supported by NSF's LIGO Laboratory, which is a major facility fully funded by the National Science Foundation.
\end{acknowledgments}

\section*{Data availability statement}
This work makes use of public GW data products \cite{ligo_scientific_collaboration_2026_19500052,ligo_scientific_collaboration_and_virgo_2021_5546676,ligo_scientific_collaboration_and_virgo_2022_6513631,ligo_scientific_collaboration_and_virgo_2023_8177023,ligo_scientific_collaboration_and_virgo_2026_20275769,ligo_scientific_collaboration_and_virgo_2026_20348005,ligo_scientific_collaboration_and_virgo_2026_20348006} and simulated sources \cite{zevin_2021_4947741}.

\bibliography{gwpop_mix}

\end{document}